\documentclass[aps,10pt,prc,tightenlines,twocolumn,twoside]{revtex4}
\usepackage{graphicx,amsmath,amssymb}
\usepackage{amsmath,amssymb}

\newcommand{\eg}{\textit{e.g.}}  

\newcommand{\ie}{\textit{i.e.}}  
\newcommand{\dd}{\mathrm{d}}
\newcommand{\lsim}{\lesssim}
\newcommand{\gsim}{\gtrsim}

\begin{document}

\title{Bottomonium Production at RHIC and LHC}

\author{L.~Grandchamp} \affiliation{Lawrence Berkeley National Laboratory,
  Berkeley, CA 94720}

\author{S.~Lumpkins} \affiliation{University of Oklahoma, Norman, OK 73019}
\affiliation{Cyclotron Institute and Physics Department, Texas A\&M
  University, College Station, TX 77843-3366}

\author{D.~Sun} \author{H.~van Hees}\author{R.~Rapp} \affiliation{Cyclotron
  Institute and Physics Department, Texas A\&M University, College Station,
  TX 77843-3366}

\date{February 10, 2006}

\begin{abstract}
  Properties of bottomonia ($\Upsilon$, $\chi_b$ and $\Upsilon'$) in the
  Quark-Gluon Plasma (QGP) are investigated by assessing inelastic
  reaction rates and their interplay with open-bottom states ($b$-quarks
  or $B$-mesons) and color-screening. The latter leads to vanishing
  quarkonium binding energies at sufficiently high temperatures (close
  to the dissolution point), which, in particular, renders standard
  gluo-dissociation, $g+\Upsilon\to b + \bar b$, inefficient due to a
  substantial reduction in final-state phase space.  This problem is
  overcome by invoking a ``quasifree'' destruction mechanism, $g,q,\bar
  q + \Upsilon \to g,q,\bar q + b + \bar b$, as previously introduced
  for charmonia.  The pertinent reaction rates are implemented into a
  kinetic theory framework to evaluate the time evolution of bottomonia
  in heavy-ion reactions at RHIC and LHC within an expanding fireball
  model. While bottom quarks are assumed to be exclusively produced in
  primordial nucleon-nucleon collisions, their thermal relaxation times
  in the QGP, which importantly figure into $\Upsilon$-formation rates,
  are estimated according to a recent Fokker-Planck treatment.
  Predictions for the centrality dependence of $\Upsilon$ production are
  given for upcoming experiments at RHIC and LHC.  At both energies,
  $\Upsilon$ {\em suppression} turns out to be the prevalent effect.
\end{abstract}

\preprint{LBNL-58479}
\maketitle


\section{Introduction}
One of the primary objectives of ultrarelativistic heavy-ion collisions
is to produce and study the Quark-Gluon Plasma (QGP), a deconfined state
of strongly interacting matter which is believed to have existed a few
microseconds after the Big Bang. Recent observations from the
Relativistic Heavy-Ion Collider (RHIC) indicate that, in a temperature
regime $T\simeq$~1-2~$T_c$ ($T_c\simeq 180~\text{MeV}$: (pseudo-)
critical temperature~\cite{Karsch:2003jg}), the QGP is not the
originally expected weakly interacting gas of quarks and gluons. These
observations could, in fact, be related to recent results from lattice
Quantum Chromodynamics (lQCD) which suggest the existence of resonance-
(or bound-state) like hadronic states up to temperatures of $T\simeq
2T_c$, both in the heavy ($Q\bar
Q$)~\cite{Umeda:2002vr,Asakawa:2003re,Datta:2003ww} and light ($q\bar
q$) quark sector~\cite{Asakawa:2002xj,Karsch:2003jg}.  For charmonium
states this opens the exciting possibility that a substantial part of
the final yield arises from (secondary) recombination of $c$ and $\bar
c$ quarks in the QGP~\cite{Thews01,GRB04} (or at the hadronization
transition~\cite{Pbm01,Goren02,GR01,An03}), as opposed to the originally
suggested signature of $J/\psi$ {\em suppression}~\cite{MS86}. The
effectiveness of recombination mechanisms in statistical and kinetic
models sensitively resides on a thermal equilibration of the charm-
($c$-) quark momentum distributions in the
QGP~\cite{Greco:2003vf,Thews:2005vj}. On the one hand, kinetic $c$-quark
equilibrium is not easily conceivable based on perturbative QCD (pQCD)
cross sections with thermal light quarks and
gluons~\cite{Svet88,Must97,HR04,Hees:2005}. On the other hand, a recent study
based on resonant rescattering via ``$D$''-meson states in the
QGP~\cite{HR04} found substantially smaller $c$-quark thermalization
times (by about a factor of $\sim$~3 compared to pQCD estimates), being
comparable to or even below the expected QGP lifetime at RHIC. Thus, the
question arises whether $J/\psi$ suppression, or rather the absence
thereof, is an appropriate QGP signature at collider energies. At this
point, the prospect of measuring $\Upsilon$ states at RHIC (and LHC)
becomes particularly valuable. Due to their large mass, $b$-quarks are
not expected to kinetically equilibrate. Together with their small
production cross sections at RHIC, the situation for bottom appears to
be reminiscent to the charm sector at SPS, in the sense that $J/\psi$
\emph{suppression} has been verified as the dominant effect, with small
contributions from regeneration~\cite{GR01,GRB04}.

The main objective of this article is to quantify predictions for
bottomonium production at collider energies under simultaneous inclusion
of both dissociation and formation reactions.  Earlier analyses of
bottomonium in heavy-ion reactions have essentially focused on
suppression mechanisms, using either the Debye-screening picture
(combined with formation time
effects)~\cite{GV97,Pal:2000zm,Goncalves:2001vn}, or gluo-dissociation
processes~\cite{Peskin:1979va,Bhanot:1979vb}, $g+\Upsilon\to b + \bar
b$~\cite{YELL03}.  In the present article we will assess the role of
backward reaction channels leading to bottomonium regeneration
(secondary production) via $b\bar{b}$ coalescence throughout the QGP
phase, as dictated by the principle of detailed balance. Importantly, we
incorporate effects of color screening via a substantial reduction of
quarkonium binding energies.  This, in turn, forces us to replace the
gluo-dissociation cross section, which becomes inefficient (and
ill-defined) for small binding energies, $\varepsilon_B\lsim
\Lambda_{\text{QCD}} \simeq T$, by ``quasifree'' scattering~\cite{GR01} of
thermal partons off the $b$ and $\bar b$ quarks in the bottomonium
state, $ g, q, \bar q + \Upsilon \to g, q, \bar q + b + \bar b$.
Despite an apparent suppression by one power of the strong coupling
constant, $\alpha_S$, in-medium reduced binding energies render
quasifree dissociation dominant over gluo-absorption due to a much
larger phase space~\cite{GR01}.
  
Our article is organized as follows. In Sec.~\ref{sec_hard} we give a
brief overview of production cross sections for both open and hidden
bottom states in hadronic collisions, which will form our baseline for
heavy-ion reactions via a scaling by the number of primordial (binary)
collisions.  In Sec.~\ref{sec_prop} we evaluate equilibrium properties
of bottomonia in the QGP (complemented by a brief motivation from recent
lQCD results in App.~\ref{app_lqcd}); we compute inelastic cross
sections (Sec.~\ref{subsec_cross}) and pertinent dissociation rates
(Sec.~\ref{subsec_life}) of $\Upsilon$'s and excited states ($\chi_b$
and $\Upsilon'$) from scattering off light partons, illustrating
limitations of the widely used gluo-dissociation and suitable
improvements. In Sec.~\ref{sec_kin} we employ a kinetic-theory framework
to study the time evolution of bottomonia in heavy-ion collisions,
characterized by an underlying rate equation introduced in
Sec.~\ref{subsec_rate-eq}.  In addition to inelastic reaction rates, the
latter requires further input consisting of (i) a space-time
(temperature) evolution of a heavy-ion reaction modelled by a thermal
fireball (App.~\ref{app_fireball}) and (ii) the equilibrium limit of
$\Upsilon$ densities (App.~\ref{app_eq-abund}).
Sec.~\ref{subsec_time-evo} contains our main results for the time
evolution of $\Upsilon$ in central Au-Au (Pb-Pb) collisions at RHIC
(LHC), with a discussion of the uncertainties related to the $\Upsilon$
equilibrium abundances relegated to appendices \ref{app_open} and
\ref{app_thermal}. The evolution and feeddown contributions from excited
bottomonia are addressed in Sec.~\ref{subsec_feed}.  In
Sec.~\ref{sec_centr} we present our predictions for the centrality
dependence of $\Upsilon$ and $\Upsilon'$ production at RHIC and LHC. We
summarize and conclude in Sec.~\ref{sec_concl} including an outlook on
future improvements of our analysis.

\section{Initial Production of Bottom(onium)}
\label{sec_hard}
The initial production of $b\bar b$ pairs and related fractions of
bottomonium states in hard (primordial) nucleon-nucleon ($N$-$N$)
collisions constitutes an essential input for our subsequent calculations.
Throughout the paper we will assume the total number of $b \bar b$ pairs in
the system to be determined by hard production alone, since, due to the
large $b$-quark mass, $m_b\gg T$, secondary production is expected to be
negligible~\cite{LMW95} (even under LHC initial conditions as considered
below, $m_b/T\gsim 4-5$).  The number of $b\bar b$ pairs at a given impact
parameter, $b$, of an $A$-$A$ collision ($A$=Au and Pb for RHIC and LHC,
respectively) can therefore be written as
\begin{equation}
N_{b\bar{b}} = \frac{\sigma_{pp\rightarrow
b\bar{b}}}{\sigma_{pp}^{\text{inelastic}}} \ N_{\text{coll}}(b) \ R_{y} \ ,
\end{equation}
where $\sigma_{pp}^{\text{inelastic}}$ denotes the total inelastic
proton-proton cross section ($\simeq$~42~mb and 78~mb for RHIC and LHC,
respectively)~\cite{PDB04}, and $N_{\text{coll}}(b)$ is the number of
primordial $N$-$N$ collisions estimated within the Glauber model.  The
inclusive cross section, $\sigma_{pp\rightarrow b\bar{b}}$, for $b\bar
b$ production in $p$-$p$ collisions is taken from recent estimates of
Refs.~\cite{Vogt02,YELL03}.  We use the values $\sigma_{pp\rightarrow
  b\bar{b}}=2.0$ $\mu$b at RHIC ($\sqrt{s_{NN}}=200$ GeV) and
$\sigma_{pp\rightarrow b\bar{b}}=160$ $\mu$b at LHC
($\sqrt{s_{NN}}=5.5$~TeV), which in the latter case includes a factor
0.8 to account for shadowing corrections. Finally, $R_{y}$=0.52~(0.29)
for RHIC (LHC) denotes the fraction of $b\bar{b}$ pairs within a given
rapidity interval ($\Delta y\simeq 3.6$ for two thermal fireballs) which
we have estimated from distributions calculated in perturbative
QCD~\cite{Vogt01}.

The primordial numbers of bottomonia are not very well known; following
Ref.~\cite{Vogt02,YELL03}, we employ $\sigma_{pp\rightarrow \Upsilon}
=$3.5~nb (152~nb) for RHIC (LHC, where we have again included a factor
of 0.8 shadowing correction).

The initial number to be used for the thermal evolution has furthermore
to account for inelastic processes in the pre-equilibrium stages. We
approximate these by the so-called nuclear absorption, which, in
principle, can be extracted from data in $p$-$A$ collisions.  Again,
this information is rather scarce. We therefore adopt the standard
approach employed for charmonia, which consists of a constant
(high-energy) nucleon-bottomonium absorption cross section,
$\sigma_{\text{nuc}}^{\text{abs}}$, coupled with a Glauber model for the
nuclear overlap function~\cite{nuc-abs}. We note that, at collider
energies, coherence effects may induce substantial deviations from the
naive collision scaling~\cite{Kopeliovich:2001ee}, typically amounting
to a significant reduction in absorption. We therefore use rather
conservative values for the absorption cross section of
$\sigma_{\text{nuc}}^{\text{abs}}=3.1$~mb at RHIC ($4.6$~mb at
LHC)~\cite{YELL03}, which, for simplicity, is assumed to be identical
for all bottomonium states.

Different values for $\sigma_{\text{nuc}}^{\text{abs}}$ will affect the
results for the centrality dependence of bottomonium production in
$A$-$A$ collisions as presented in Sec.~\ref{sec_centr}. On the one
hand, the number of ``directly'' produced bottomonia, defined as the
primordial number subjected to suppression only, $N_{\rm dir}=N_{\rm
  prim} \ S_{\rm nuc} \ S_{\rm QGP}$, depends on the nuclear absorption
cross section through the nuclear suppression factor, $S_{\text{nuc}}$
(schematically, $S_\text{nuc} \approx
\exp[-\rho_N~\sigma_{\text{nuc}}^{\text{abs}}~L(b)]$ with an effective
nuclear density, $\rho_N\simeq$0.14~fm$^{-3}$, and impact parameter
dependent path length, $L(b)$).  On the other hand, the number of
regenerated bottomonia is practically independent of
$\sigma_{\text{nuc}}^{\text{abs}}$.

\section{Upsilon Properties in the QGP}
\label{sec_prop}

In this section we will assess
inelastic cross sections and dissociation rates of the $\Upsilon$
and its excited states in the QGP. 
Our main objective is a combined and consistent treatment of
color screening of the heavy-quark potential~\cite{MS86} and 
inelastic reaction channels. In analogy to the present situation 
for (ground state) charmonia, and motivated by lQCD results,
we assume the bottomonium masses to be constant. The hadronic phase 
will be neglected altogether throughout this paper. 
Whenever possible, we will try to motivate necessary assumptions
by findings of recent lQCD calculations (see App.~\ref{app_lqcd}
for a brief survey containing more details). 

\subsection{Dissociation Cross Sections}
\label{subsec_cross}
\begin{figure}[!tbp]
  \includegraphics[width=0.9\linewidth,clip=]{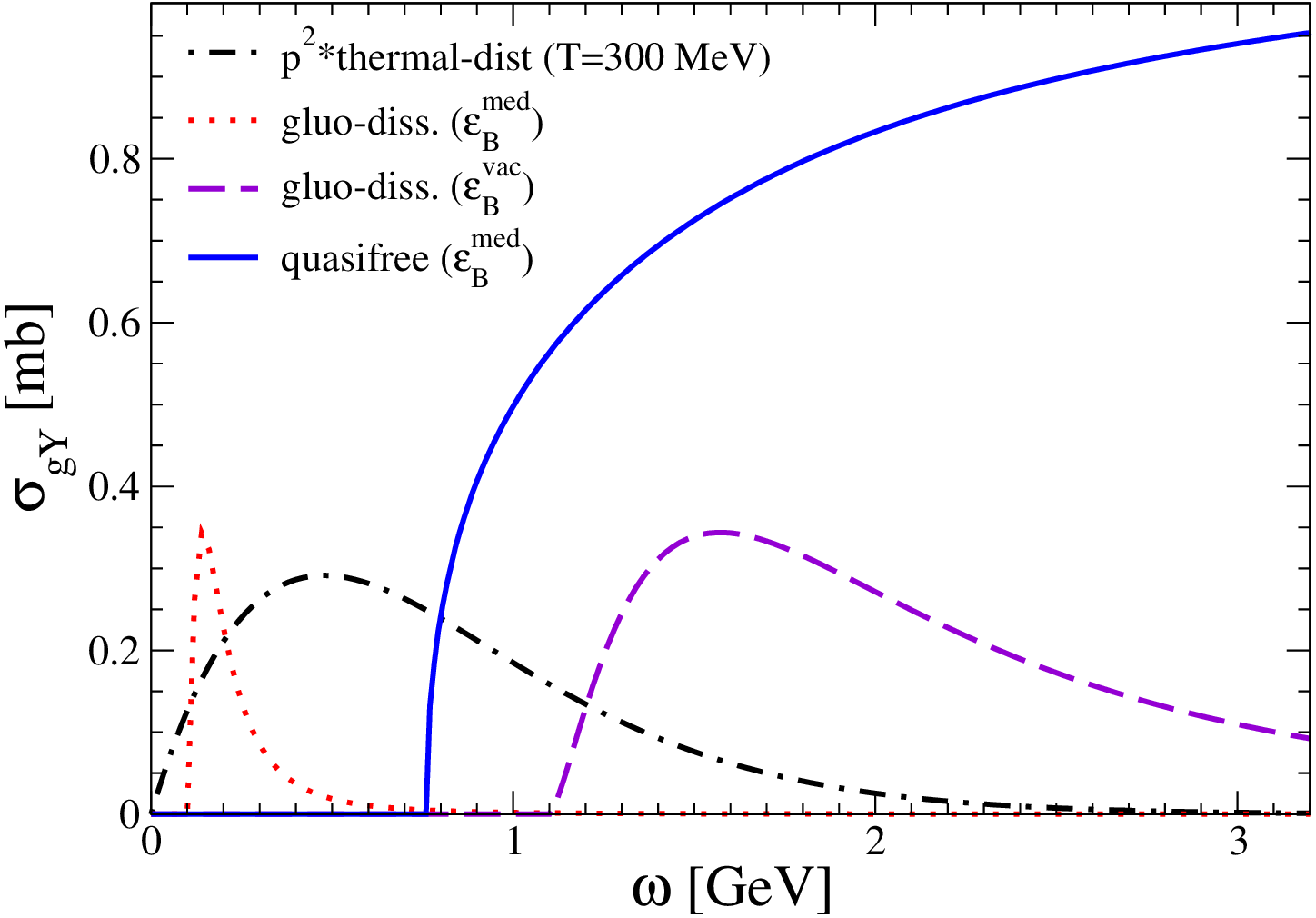}
  \caption{(Color online) Gluo-dissociation (with vacuum binding energy:
    dashed line, or with in-medium reduced binding energy: dotted line)
    and quasifree (solid line) cross section overlap with a thermal
    gluon distribution at $T=300$~MeV (dash-dotted line).}
  \label{fig_sig-diss}
\end{figure}

The most commonly used process to evaluate the bottomonium (or 
charmonium) dissociation cross section, 
$\sigma_{Y}^{\text{diss}}$ ($Y$=$\Upsilon$, $\chi_b$, $\Upsilon'$),  
on partons is the QCD analogue of 
photo-dissociation, $g + Y \rightarrow b + 
\bar{b}$~\cite{Peskin:1979va,Bhanot:1979vb}.
In Fig.~\ref{fig_sig-diss}, the dashed line represents
$\sigma_{\Upsilon}^{\text{diss}}$ 
for the ground state $\Upsilon(1s)$ with vacuum binding energy, 
$\varepsilon_B^{\text{vac}}\simeq 1.1$ GeV (defined with respect to 
the $B\bar B$ threshold, $2m_B$), 
as a function of gluon energy $\omega$ in the $\Upsilon$-rest frame.
It is peaked around $\omega \simeq 1.5 \, \varepsilon_B \simeq 1.5$~GeV, 
which for moderate plasma temperatures, $T\simeq$~300~MeV, is 
significantly above average thermal gluon energies, $\omega=3T$; 
in other words, 
the maximum of the cross section has rather little overlap with
the phase-space weighted thermal gluon distribution, 
$p^2 f^g(\omega;T)$ (dash-dotted line in Fig.~\ref{fig_sig-diss}).    
While an increase in $T$ and a reduction in binding energy (as
expected from color-screening) initially improves the
overlap, a small value of $\varepsilon_B$  (toward/close to the
dissolution point) eventually leads to a very narrow energy interval
where the cross section is active (dotted line in Fig.~\ref{fig_sig-diss})
and thus renders gluo-dissociation increasingly inefficient. 
Formally this is due to the fact that for a loosely bound (large-size) 
$b$-$\bar b$ system, the absorption of a hard gluon essentially occurs 
on a single $b$-quark, in which case energy-momentum conservation cannot 
be satisfied, and higher orders in the multipole expansion have to
be included. This problem is even more severe for the less bound 
(excited) bottomonia. 

To implement effects of color-screening we follow the approach suggested
in Ref.~\cite{GR01,GR02} in the context of charmonia: for small binding
energies we approximate $Y$-states as two comoving $b$ and $\bar{b}$
quarks, which individually interact with surrounding thermal partons via
lowest-order (quasifree) pQCD scattering~\cite{Comb79}, $g (q,\bar{q}) +
b \to g (q,\bar{q}) + b$ (and likewise for $\bar b$).  The reaction is
considered to lead to $Y$ breakup, $g (q,\bar{q}) + \Upsilon \rightarrow
g(q,\bar{q}) + b +\bar{b}$, if the residual binding energy can be
overcome (and conserve overall four-momentum). Note that in this
approach also thermal (anti-) quarks contribute to bottomonium
dissociation via $t$-channel gluon-exchange.  The $t$-channel
singularities are regularized by a gluon Debye-mass, $\mu_D\sim gT$,
with $\alpha_S\simeq 0.26$ as fixed in our analysis for
charmonia~\cite{GRB04} (note that the $t$-channel gluon-propagator
compensates one power in $\alpha_S$, which formally renders the
quasifree cross section of the same order as gluo-dissociation).  The
result for the quasifree dissociation cross section of $\Upsilon(1s)$
with in-medium binding energy (solid line in Fig.~\ref{fig_sig-diss})
shows a much improved overlap with thermal distributions due to its
monotonous rise with gluon energy. The threshold for the result shown is
due to a combination of binding energy and thermal parton masses; for
massive partons, necessary to reproduce the QGP equation of state, the
efficiency of gluo-dissociation is even more suppressed if
$\varepsilon_B < m_g \sim gT$. In addition, the quasifree cross section
is theoretically better controlled for small values of the binding
energy than gluo-dissociation since its main contribution arises from
gluons with relatively large (thermal) energies, $\omega\simeq 3T$.  Its
applicability therefore also encompasses excited bottomonia.

\subsection{Lifetimes}
\label{subsec_life}

To evaluate bottomonium suppression in the plasma, the inelastic cross
section, $\sigma_{Y}^{\text{diss}}$, is converted into a dissociation
rate $\Gamma_{Y}$ (or inverse lifetime $\tau_{Y}^{-1}$) by convoluting
it with thermal distributions, $f_{g,q}$, of quarks and gluons,
\begin{equation}
  \label{eq:1}
  \Gamma_{Y}\equiv\tau_{Y}^{-1}=\int
  \frac{d^3k}{(2\pi)^3} \, f_{q,g}(\omega_k,T) \, v_{\text{rel}} \,
  \sigma_{Y}^{\text{diss}}(s) \ . 
\end{equation}
Here, $s=(q+k)^2$ denotes the total center-of-mass (CM) energy squared
of the collision of a $Y$ at rest ($q=(m_Y,\vec 0)$) with a thermal
on-shell parton of four-momentum $k$. We repeat that throughout this
paper we use the vacuum masses for the bottomonia (for simplicity we do
not distinguish the three spin states of the $\chi_{b}$ but use their
average mass).

\begin{figure}[!tbp]
    \includegraphics[width=0.9\linewidth,clip=]{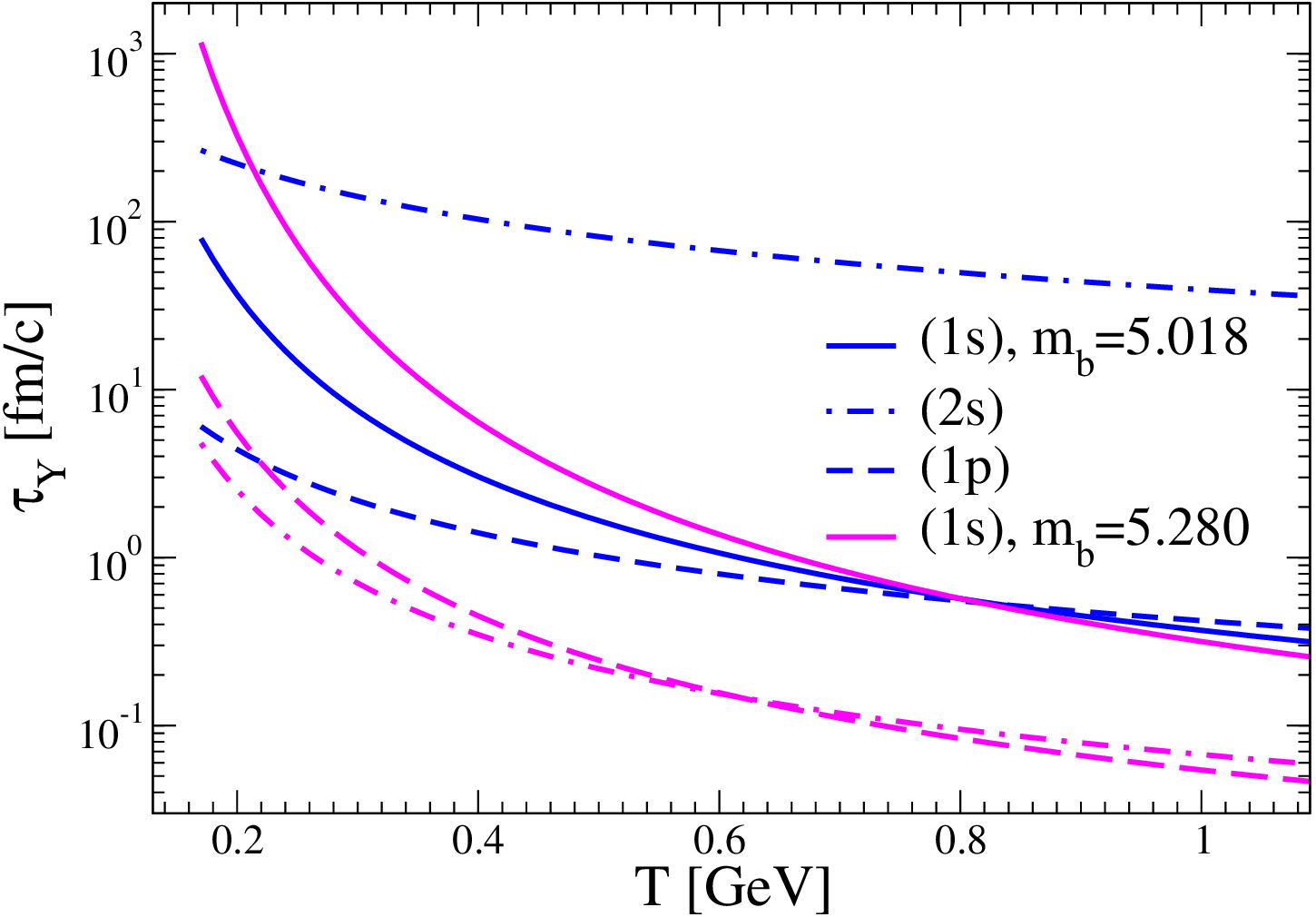}
    \caption{(Color online) $Y$ lifetimes vs. temperature in a QGP based
      on the gluo-dissociation process for two ($T$-independent) values
      of the bottom-quark mass and pertinent binding energies according
      to Eq.~(\ref{mb}).  Solid lines: $\Upsilon$, dash-dotted lines:
      $\Upsilon'$, dashed lines: $\chi_b$. The light- (dark-) colored
      set of curves corresponds to $m_b=5.280$~GeV ($m_b=5.018$~GeV).}
  \label{fig:rates-gluo-vac}
\end{figure}

Let us start by discussing the lifetimes arising from gluo-dissociation
using constant (temperature independent) $Y$ binding energies for two
different values of the open-bottom threshold (2$m_b$),
cf.~Fig.~\ref{fig:rates-gluo-vac}.
For the larger $m_b=5.280$~GeV (light colored curves), corresponding 
to the lightest $B$-meson, the lifetimes are rather well-behaved at
temperatures relevant for RHIC ($T\lsim 400$~MeV), \ie, rather large 
for the $\Upsilon$ (solid line, around 10~fm/c and above, which is
significantly longer than the expected QGP lifetime), much smaller 
for the $\chi_b$ (dashed line, 1~fm/c and below), and still slightly 
smaller for the less bound $\Upsilon'$ (dash-dotted line).   
However, for temperatures in excess of $\sim$500~MeV, the $\Upsilon'$ 
lifetime becomes larger than the one of the $\chi_b$, a first 
indication of artifacts induced by the gluo-dissociation cross section
for small binding energies as discussed in the previous section. 
This trend becomes more pronounced upon reducing the bottom-quark
mass to  $m_b=5.018$~GeV (dark-colored set of lines in
Fig.~\ref{fig:rates-gluo-vac}), implying 
$\varepsilon_B^{\Upsilon'}=13$~MeV and $\varepsilon_B^{\chi_b}=145$~MeV. 
In this scenario, the $\Upsilon'$
lifetime is roughly two orders of magnitude larger than the $\Upsilon$
and $\chi_b$ lifetimes, which is obviously unrealistic and clearly
indicates deficiencies of the gluo-dissociation
cross section.

\begin{figure}[!tb]
\includegraphics[width=0.9\linewidth,clip=]{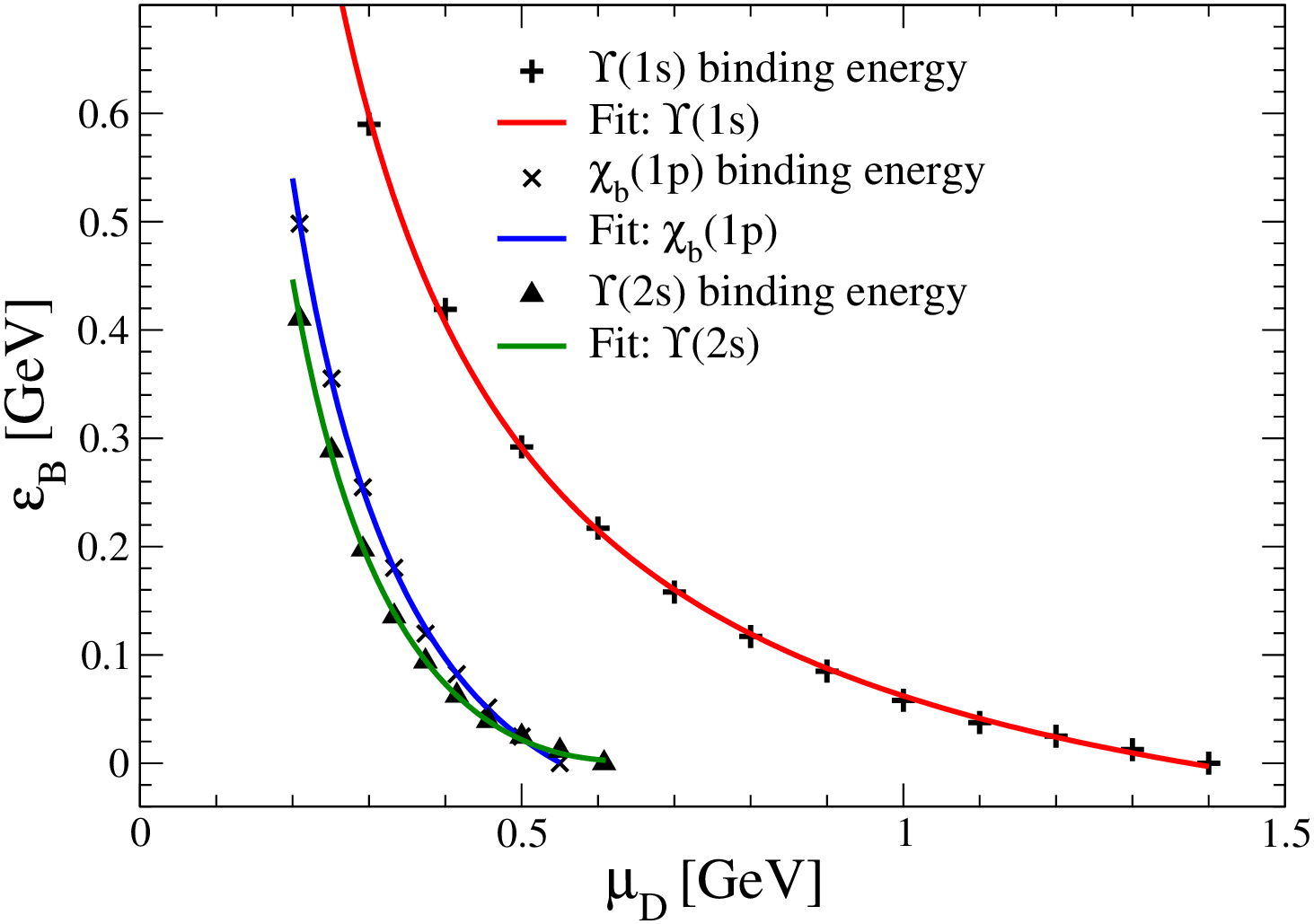}
\caption{(Color online) The dependence of $Y$ binding energies (symbols)
  on gluon screening (Debye) mass according to solutions of the
  Schr\"odinger equation with a screened Cornell potential~\cite{KMS88}.
  The lines indicate fits used for the numerical implementation in our
  calculations. The identification with temperature is provided by the
  perturbative expression for the Debye mass (see text), amounting to,
  \eg, $\mu_D(T_c)\simeq$~325~MeV.}
  \label{fig:Ediss}
\end{figure}

\begin{figure}[!tb]
\includegraphics[width=0.9\linewidth,clip=]{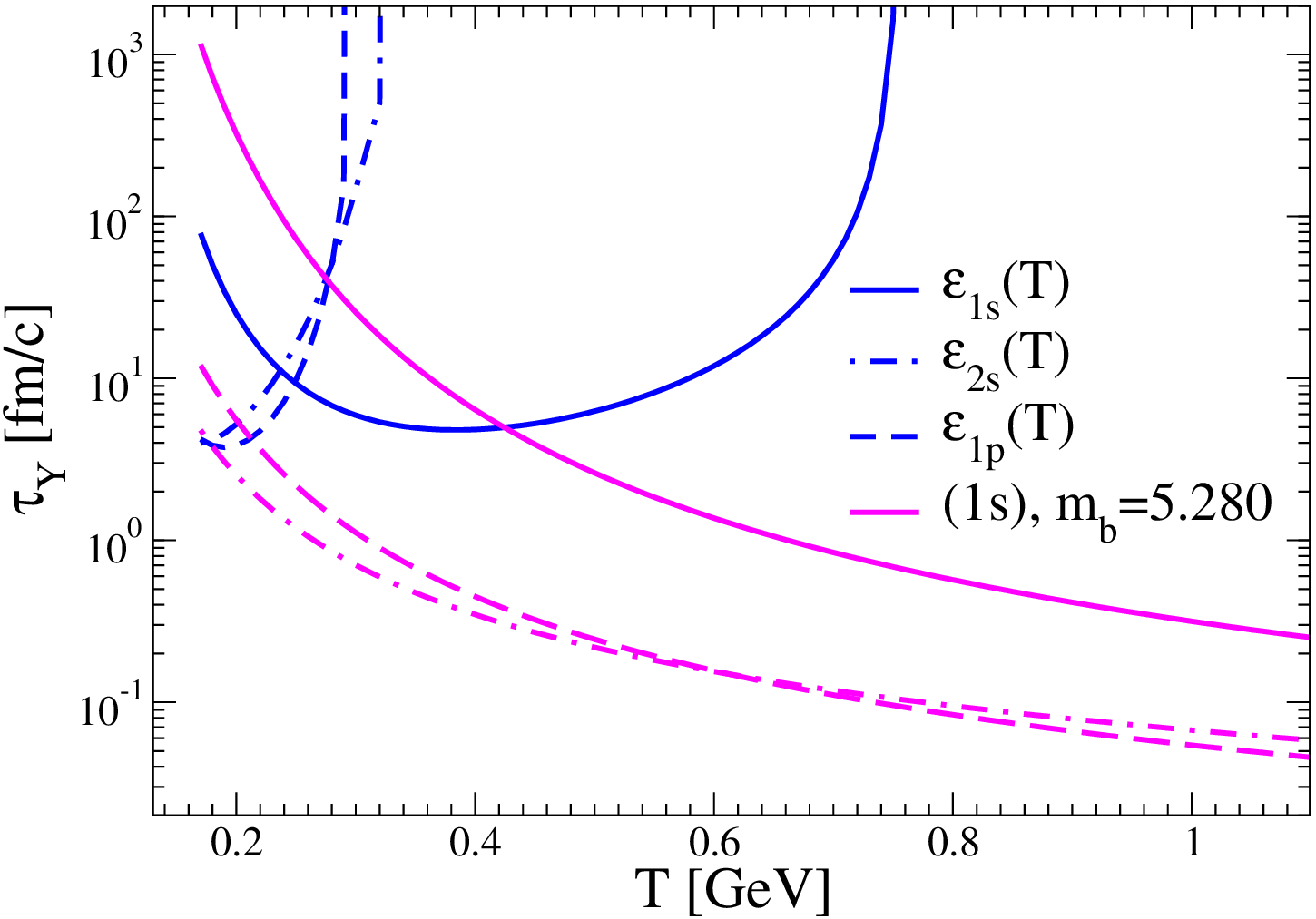}
\caption{(Color online) $Y$ lifetimes vs. temperature in a QGP based on
  gluo-dissociation with an in-medium bottom-quark mass according to
  in-medium reduced binding energies as computed in Ref.~\cite{KMS88}
  (dark-colored curves).  Solid lines: $\Upsilon$, dash-dotted lines:
  $\Upsilon'$, dashed lines: $\chi_b$.  The light colored curves are
  identical to Fig.~\ref{fig:rates-gluo-vac}, \ie, for constant binding
  energy with $m_b=5.280$~GeV.}
  \label{fig:rates-gluo-med}
\end{figure}

As a more systematic way to include the effects of in-medium 
heavy-quark potentials in our calculations we make use of the results 
of Ref.~\cite{KMS88} obtained from a Schr\"odinger 
equation using a Debye-screened Cornell potential.
Assuming a Debye mass of the form suggested by pQCD, 
$\mu_D=gT$ (with $g\simeq 1.8$ as used in the
quasifree cross section), the $\Upsilon$ ($\chi_b$, $\Upsilon$') 
binding energy is found to be reduced to about $\sim$550~MeV 
(150-200~MeV) at $T_c$, reaching zero at about 
4.3~$T_c$ ($\sim$1.8~$T_c$), cf.~Fig.~\ref{fig:Ediss}. These values
are in reasonable agreement with solutions to a Schr\"odinger 
equation based on recent (unquenched) lQCD potentials (except
close to $T_c$)~\cite{Alberico:2005xw}.
Since the screening close to $T_c$ might well be less pronounced, our
in-medium scenario should be considered as an upper limit of a reduction
in binding energies. The pertinent effects in connection with the
gluo-dissociation cross section are illustrated by the dark-colored
curves in Fig.~\ref{fig:rates-gluo-med}, (the light-colored curves are
identical to Fig.~\ref{fig:rates-gluo-vac}).
Whereas close to $T_c$ the ordering is still reasonable, the strong
reduction of $\varepsilon_B$ with increasing $T$ rapidly induces a very
unrealistic behavior in terms of a reduction in the dissociation rate
(increase in lifetime).  As discussed above, this artifact is induced by
a large decrease in available phase space for the gluo-dissociation
cross section which renders bottomonium dissociation a very inefficient
reaction as the temperature increases.

\begin{figure}[!tbp]
  \centering
  \includegraphics[width=0.9\linewidth,clip=]{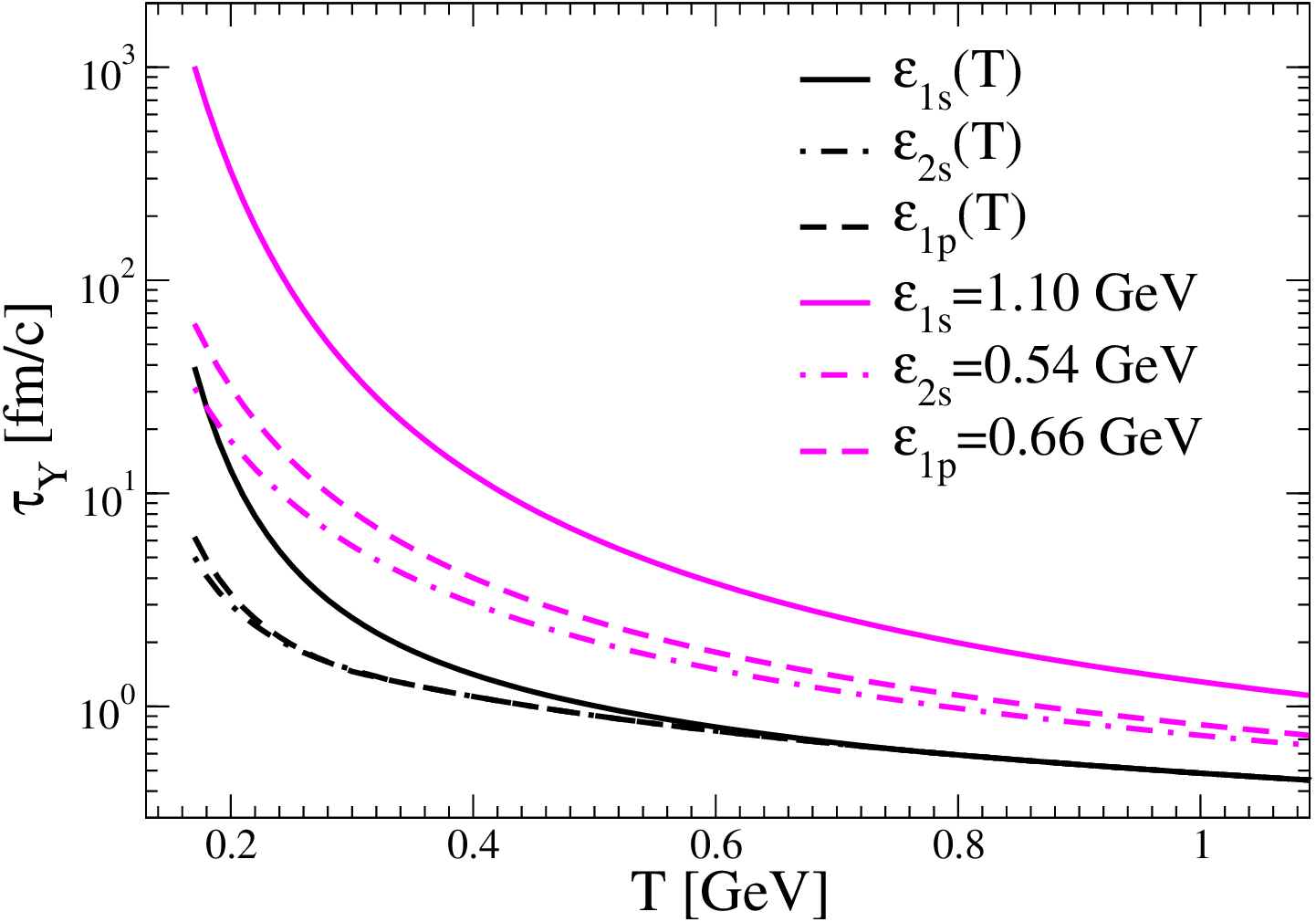}
  \caption{(Color online) Y lifetimes computed from quasifree
    dissociation using constant (light set of curves) and in-medium
    (dark set of curves) binding energies for the $\Upsilon$ (solid
    line), $\Upsilon'$ (dash-dotted line) and the $\chi_b$ (dashed
    line). }
  \label{fig:rates-quasifree}
\end{figure}

As indicated in the previous section the (presumably unphysical)
decrease in the reaction rates with increasing $T$ can be remedied by
introducing the quasifree dissociation processes which are not limited
by phase space and therefore apply to excited states as well.  The
results are summarized in Fig.~\ref{fig:rates-quasifree}, where the
corresponding $Y$ lifetimes are plotted as a function of temperature.
The light-colored lines represent vacuum binding energies with an
open-bottom mass of $m_b=5.280$~GeV. The resulting lifetimes are
somewhat larger than the ones using gluo-dissociation (light-colored
lines in Fig.~\ref{fig:rates-gluo-vac}), but exhibit the (from the size
and binding energies) expected hierarchy,
$\tau_{\Upsilon'}<\tau_{\chi_b}<\tau_{\Upsilon}$, at all temperatures.
The real virtue of the quasifree approach manifests itself in a
(presumably more realistic) calculation using in-medium ($T$-dependent)
binding energies corresponding to the dark-colored lines. Here, smaller
$\varepsilon_B$ facilitate bottomonium breakup, and the respective
dissociation rates are always larger than the ones using the vacuum
values for $\varepsilon_B$.  In comparison to the gluo-dissociation
results, two further remarks are in order: (i) At the highest
temperatures the $Y$ lifetimes are larger than the ones calculated with
gluo-dissociation using vacuum binding energies. The main reason for
this is that the strong coupling constant, $\alpha_S$, inherent in the
$b$-$\bar b$ Cornell potential necessary to reproduce bottomonium
spectroscopy (which also enters the dissociation cross sections), is
substantially larger ($\sim$0.5) than in the quasifree calculation
($\sim$0.26). After all, it is the Debye-screening of the potential
which is related to an effectively reduced $\alpha_S$ in the medium.
(ii) At temperatures $T\lsim 300$(200)~MeV the gluo-dissociation rates
for $\Upsilon$ ($\Upsilon'$, $\chi_b$) using in-medium binding energies
are not much smaller than the ones obtained with quasifree destruction.
Therefore, both contributions to the rate should, in principle, be
accounted for (at higher $T$, gluo-dissociation with in-medium
$\varepsilon_B$ is negligible).  However, since we have fixed $\alpha_S$
in the charmonium sector using the quasifree process alone to reproduce
$J/\psi$ data in Pb-Pb collisions at the SPS, we will refrain from
adding the two contributions. In addition, as we will see below, most of
the bottomonium suppression at RHIC and LHC occurs at temperatures above
$\sim$250~MeV.

Unless otherwise specified, in the following we will use the 
bottomonium lifetimes pertaining to the quasifree mechanism with 
in-medium binding energies as our default scenario (we recall that
this should provide an upper estimate of the screening effects).
For comparison purposes with existing literature, we will 
also discuss results using the gluo-dissociation reaction alone
with vacuum binding energies.

\section{Kinetic Theory Approach}
\label{sec_kin}
To calculate the time evolution and final abundances of bottomonia 
in a high-energy heavy-ion collision, we employ a kinetic  
rate-equation approach, as has recently been applied to charmonium
production~\cite{Thews01,GRB04}. On the one hand, such an approach 
has the appealing feature that it makes explicit contact to 
equilibrium properties of  quarkonia~\cite{GRB04}, as discussed in 
the previous section. On the other hand, it does not provide
an accurate treatment for processes occurring far from equilibrium,
in which case explicit transport calculations are 
preferable~\cite{Zhang:2002ug,Bratkovskaya:2003ux}.  

\subsection{Rate Equation}
\label{subsec_rate-eq}
In the limit that bottom quarks (and/or hadronic open-bottom states 
which might survive up to $T\simeq$~2$T_c$ or so) are in thermal 
equilibrium with the bulk of the system (light partons), the time 
dependence of the $Y$ number, $N_Y$,
obeys a simplified rate equation given by~\cite{RG03}
\begin{equation}
\label{rate-eq}
\frac{\dd N_{Y}}{\dd
  \tau}=-\Gamma_{Y} \ (N_{Y}-N_{Y}^{\text{eq}})
 \ ,
\end{equation}
where spatial (temperature) gradients have been neglected. Here,
$\Gamma_{Y}$ denotes the $Y$-dissociation rate discussed in
Sec.~\ref{subsec_life} above and $N_{Y}^{\text{eq}}$ the equilibrium
number.  The first term on the right-hand side of Eq.~(\ref{rate-eq})
describes the loss of bottomonia due to dissociation, whereas the second
term takes into account regeneration according to the principle of
detailed balance,
\begin{equation}
\label{reaction}
X_1 + Y \leftrightarrow X_2 + b+\bar{b} \ ,
\end{equation} 
where $X_1$ and $X_2$ are light-quark or gluon states.
Obviously, the backward channel amounts
to a recombination (or coalescence) of a $b$- and $\bar{b}$-quark
within the QGP. From Eq.~\eqref{rate-eq} it is clear that
the regeneration contribution to the $Y$ yield is negligible if
either the equilibrium number is small compared to the actual number,
$N_{Y}^{\text{eq}} \ll N_{Y}$ at all times, or the
dissociation/recombination processes shut off altogether,
\ie, the reaction rate $\Gamma_{\Upsilon}$ is very small.

Eq.~(\ref{rate-eq}) is amenable to an explicit solution which reads
\begin{equation}
\label{sol}
N_{Y}(t)= S_{\rm QGP}(t) 
\left [ N_Y^0 + \int_0^t \dd \tau \frac{\Gamma_{Y}(\tau)
    N^{\text{eq}}_Y(\tau)}{S_{\rm QGP}(\tau)} \right],
\end{equation}
where 
\begin{equation}
\label{re.4}
S_{\rm QGP}(t)=\exp \left [-\int_0^{t} \dd \tau \Gamma_{\Upsilon}(\tau) 
\right ]  
\end{equation}
is the QGP suppression factor representing the probability that an $Y$,
present at $t=0$, has survived at time $t$, and $N_Y^0$ is the number of
primordially produced $Y$'s after pre-equilibrium (nuclear) absorption.

The solution to the rate Eq.~(\ref{rate-eq}) requires the time
dependence of the temperature, $T(t)$. To this end, we describe the
space-time evolution of the system with a simple thermal fireball
expansion~\cite{Rap01}. It is constructed to reproduce the main features
(timescales and transverse flow velocities) of hydrodynamical
simulations of A-A collisions~\cite{KR03}, and has been applied before
to thermal dilepton~\cite{Rapp:2005} and
charm(onium)~\cite{GR02,Hees:2005} production at SPS and RHIC energies,
cf.~App.~\ref{app_fireball} for more details.
\begin{figure}[!t]
  \centering \includegraphics[width=0.9\linewidth]{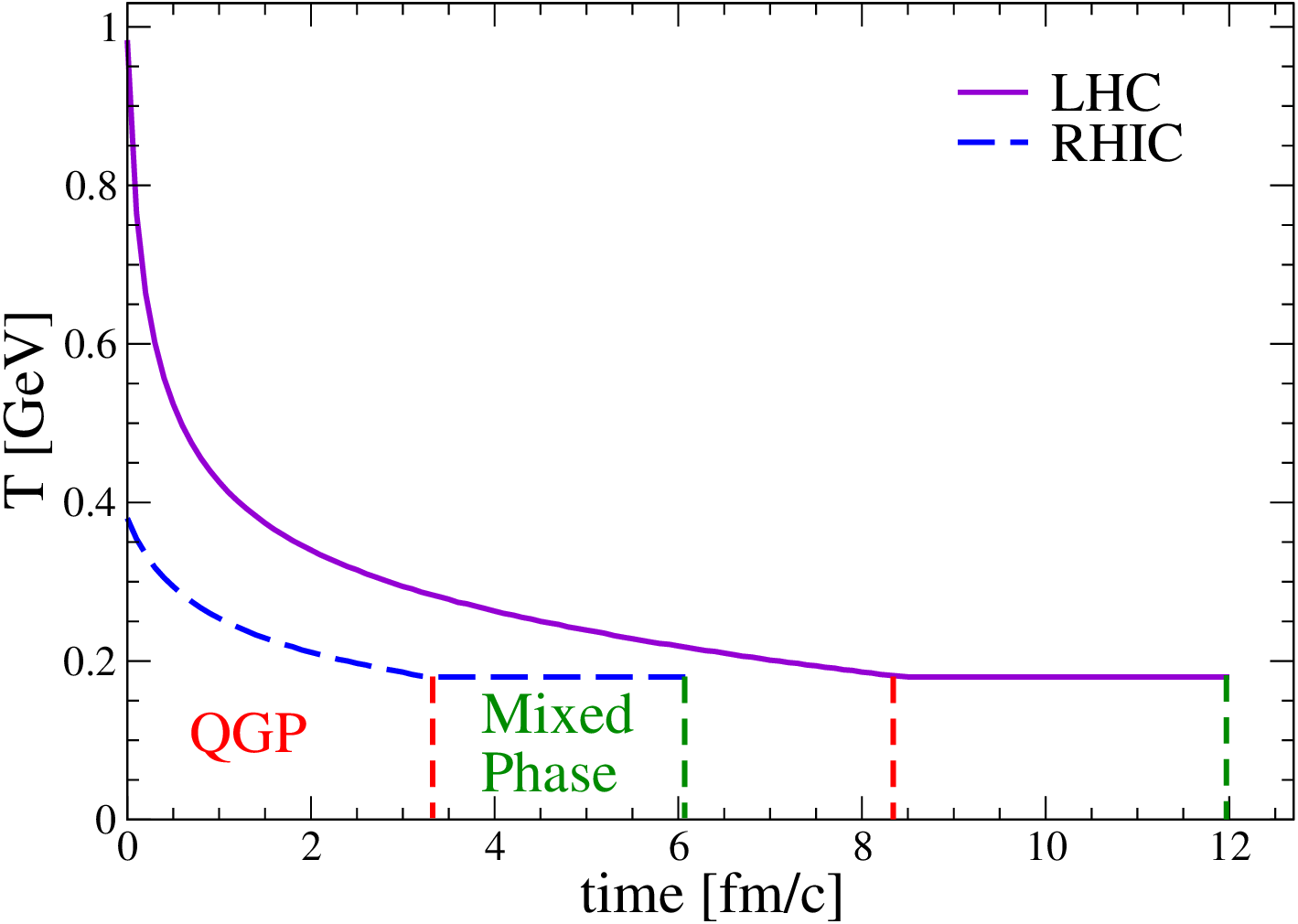}
  \caption{(Color online) Time evolution of temperature within our
    expanding fireball model for a massive parton gas in central $A$-$A$
    collisions at RHIC (dashed line) and LHC (solid line). The use of
    massless partons (with $N_f=2.5$) affects the evolution rather
    moderately.}
\label{fig_temp-profil}
\end{figure}
The resulting temperature profiles are summarized in 
Fig.~\ref{fig_temp-profil}:
the pure QGP phase lasts for about 3~fm/c at RHIC
with initial temperatures, $T_0$, close to 400~MeV, while its duration
is $\sim$8~fm/c at LHC with $T_0=T(\tau_0)\simeq 1$~GeV.

\begin{figure}[!tbp]
   \centering
\includegraphics[width=0.9\linewidth,clip=]{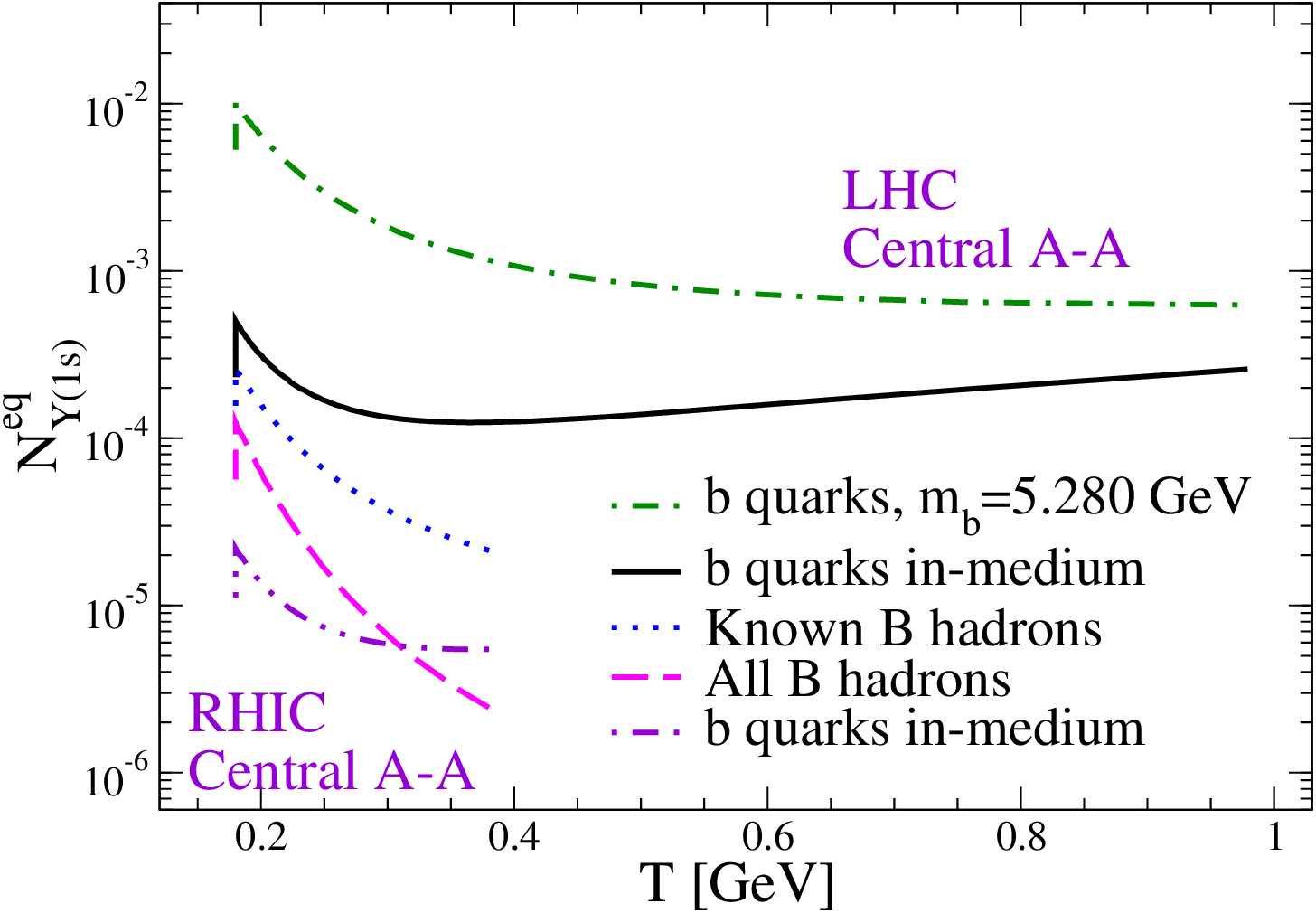} 
\caption{(Color online) $\Upsilon$(1s)-equilibrium abundances for
  central $A$-$A$ collisions at RHIC (lower 3 curves) and LHC (upper 2
  curves), according to different scenarios for open-bottom states in
  the QGP (see text and App.~\ref{subapp_open}).}
   \label{fig_open}
 \end{figure}

The final ingredient needed for the solution of the rate
Eq.~(\ref{rate-eq}) is the equilibrium number, $N_Y^{\text{eq}}$, of $Y$
states, governing their regeneration contribution throughout the
evolution. As elaborated in App.~\ref{app_eq-abund}, for a fixed number
of $b\bar b$ pairs in the system, $N_Y^{\text{eq}}$ depends on the
spectrum of open-bottom states in the QGP. We consider the following
possibilities: \\
(i) $b$-quarks with $m_b=5.280$ GeV,\\
(ii) $b$-quarks with in-medium mass $m_b(T)$ following from the 
relation 
\begin{equation}
 2m_b(T) = m_{Y} + \varepsilon_B^{Y}(T) \
\label{mb}
\end{equation}
with constant (free) $Y$-masses and binding energies according
to Fig.~\ref{fig:Ediss},\\
(iii) $B$-hadrons as listed by the particle data group~\cite{PDB04},
and\\
(iv) open-bottom hadrons as extrapolated from known charmed hadrons.\\
The general trend is that a higher degeneracy and a smaller mass of 
available open-bottom states favors $b$ and $\bar b$ quarks to 
reside in the former and thus entails a smaller equilibrium
abundance of bottomonia, as reflected in Fig.~\ref{fig_open} 
(cf.~also App.~\ref{subapp_open}). 

 So far our discussion has been residing on the assumption that
 open-bottom states have momentum distributions in thermal equilibrium
 with the bulk matter. However, due to the rather short duration of a
 few fm/c of the QGP phase in URHIC's, and a $b$-quark mass that exceeds
 expected early temperatures by a large factor, this is presumably not a
 realistic assertion~\cite{HR04,Hees:2005}. Following a suggestion made
 in Ref.~\cite{GR02}, all our results displayed in the main text below
 therefore contain a schematic correction to account for incomplete
 $b$-quark equilibration, see App.~\ref{subapp_off-equil}, with the
 pertinent thermal relaxation times taken from
 Refs.~\cite{HR04,Hees:2005} which in turn provide a reasonable
 description of recent RHIC data for single-electron spectra (attributed
 to $D$- and $B$-meson
 decays)~\cite{PHENIXv2,gag05}.
 The associated numerical uncertainty is illustrated by considering the
 thermal-equilibrium limit in App.~\ref{app_thermal}.

\subsection{Time Evolution of $\Upsilon$ in Central $A$-$A$}
\label{subsec_time-evo}
With all ingredients in place we proceed to evaluate the time evolution
of the number of $\Upsilon$ states in central Au-Au (Pb-Pb) collisions
by convoluting the rate equation \eqref{rate-eq} over the thermal
fireball expansion. As a reference scenario commonly used in the
literature, we first consider $\Upsilon$ interactions using the
gluo-dissociation process with vacuum binding energies. We then turn to
our default scenario corresponding to the quasifree mechanism with
in-medium binding energies.

\subsubsection{Gluo-Dissociation}
\label{subsubsec_gluodiss}

\begin{figure}[!tb]
  \includegraphics[width=0.9\linewidth,clip=]{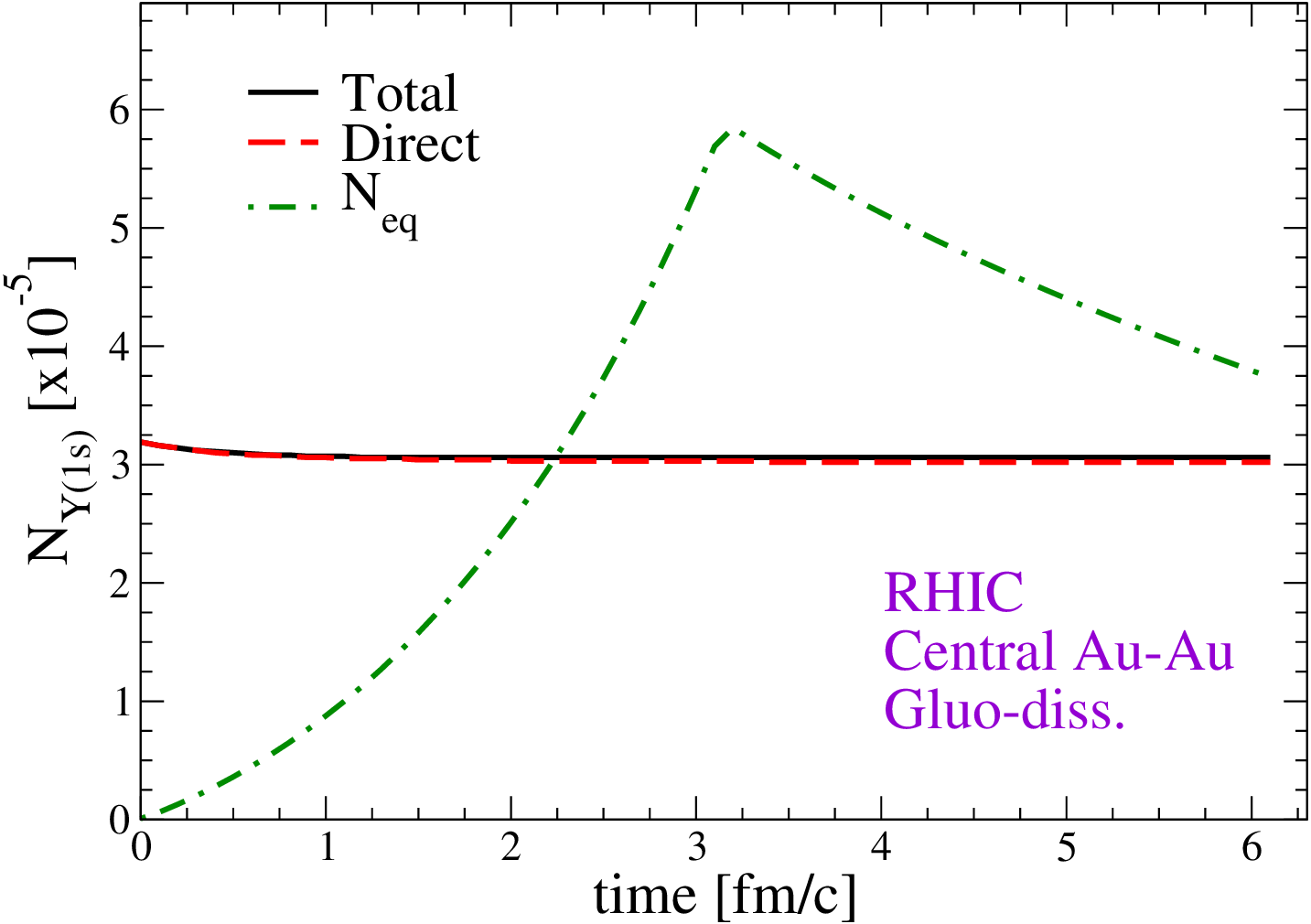}
  \caption{(Color online) $\Upsilon$(1s) abundance as a function of time
    for central ($b=1$~fm) Au-Au collisions at RHIC using the
    gluo-dissociation process (without feeddown from excited states).
    The solid line includes $\Upsilon$ suppression and regeneration and
    is almost indistinguishable from the dashed line corresponding to
    $\Upsilon$ suppression only. The dash-dotted line represents the
    $\Upsilon$-equilibrium number which governs the regeneration term in
    Eq.~\eqref{reaction}.}
  \label{fig:gluo-evol-std-rhic}
\end{figure}

\begin{figure}[!tbp]
  \includegraphics[width=0.9\linewidth,clip=]{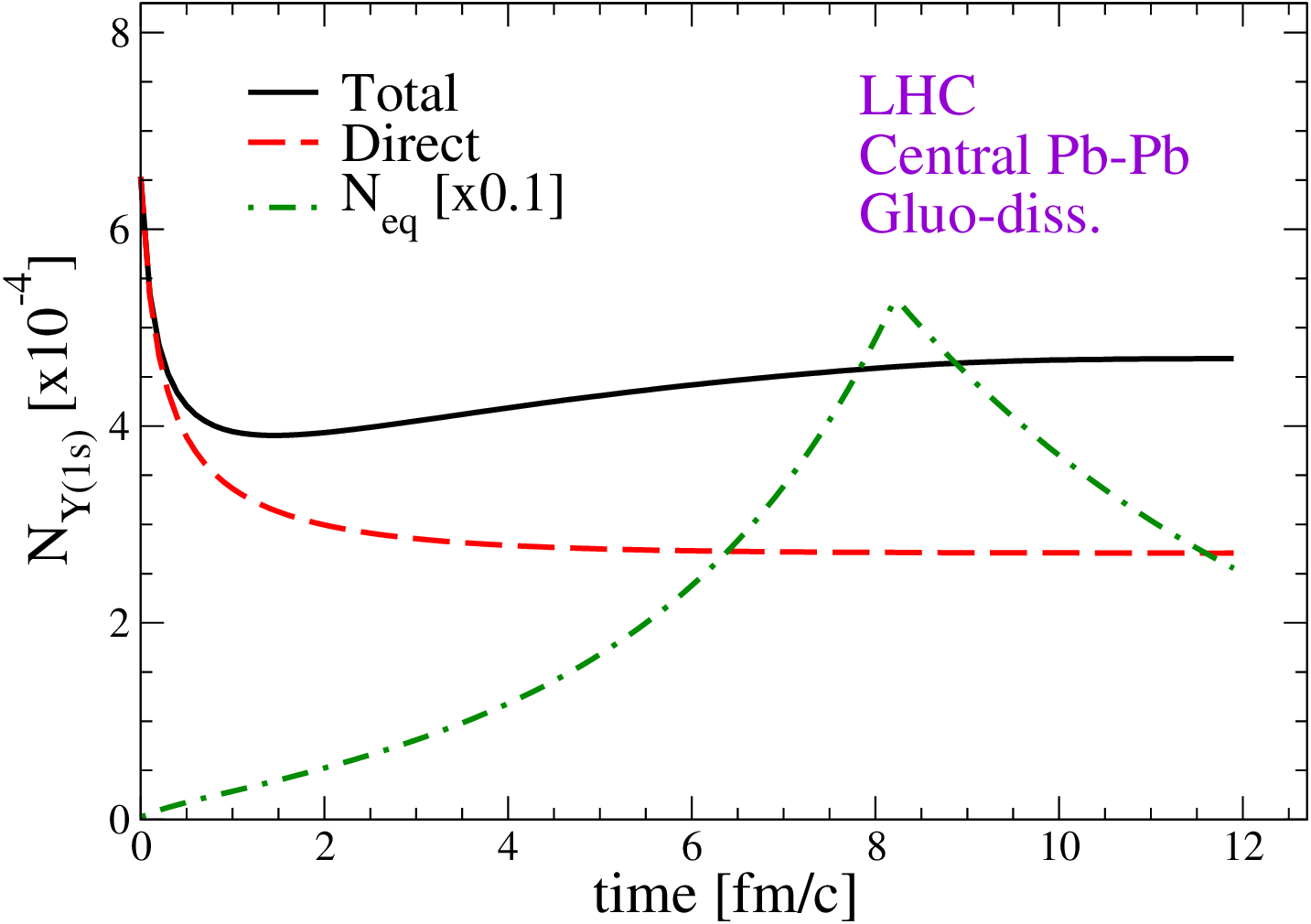}
  \caption{(Color online) $\Upsilon$(1s) abundance as a function of time
    for central ($b=1$~fm) Pb-Pb collisions at LHC using the
    gluo-dissociation suppression mechanism (without feeddown from
    higher bottomonium states). The solid line, including both
    suppression and regeneration, indicates substantial secondary
    $\Upsilon$ production when compared to the dashed line (no gain
    term), facilitated by rather large $\Upsilon$-equilibrium numbers
    (dash-dotted line) and reaction rates.}
  \label{fig:gluo-evol-std-lhc}
\end{figure}

As elaborated in Sec.~\ref{subsec_life}, in the gluo-dissociation
scenario we are limited to using vacuum binding energies which we define
relative to the hadronic open-bottom threshold. This amounts to equating
the bottom-quark mass, $m_b$, to the mass of the lightest bottom hadron,
\textit{i.e.}, $m_b = m_B = 5.280$ GeV. This rather large mass entails
comparatively large $\Upsilon$-equilibrium numbers (cf.~dash-dotted
lines in Fig.~\ref{fig:gluo-evol-std-rhic} (RHIC) and
\ref{fig:gluo-evol-std-lhc} (LHC)), which for both RHIC and LHC
significantly exceed primordial production in the later stages of the
evolution.

Nevertheless, at RHIC the effect of regeneration is essentially 
irrelevant (full vs. dashed line in Fig.~\ref{fig:gluo-evol-std-rhic})  
due to small reaction rates at the relevant temperatures
of $T<400$~MeV, cf.~Fig.~\ref{fig:rates-gluo-vac}. 

The situation is different at LHC (Fig.~\ref{fig:gluo-evol-std-lhc})
where secondary production is found to be at the 40\% level (dashed vs.
full line Fig.~\ref{fig:gluo-evol-std-lhc}), due to a combination of a
longer plasma lifetime (facilitating the build-up of a larger
``equilibrium'' limit due to thermal relaxation of $b$-quarks) as well
as the higher initial temperatures where the inelastic reaction rates
are much increased (cf.~Fig.~\ref{fig:rates-gluo-vac}).  However, within
the gluo-dissociation framework the importance of regeneration at LHC
crucially hinges on the large $\Upsilon$-equilibrium abundance induced
by the large mass of the bottom quark, $m_b=5.280$~GeV, which in turn
was used to consistently implement the vacuum binding energy of the
$\Upsilon$.  This behavior will be modified if in-medium binding
energies are employed, as we will see in the following subsection.

\subsubsection{Quasifree Dissociation}
\label{subsubsec_quasifree}
As discussed above, we use the quasifree approach to inelastic 
reaction rates to implement in-medium reduced $\Upsilon$ binding 
energies following the estimates within a screened heavy-quark
(Cornell) potential of Ref.~\cite{KMS88}. Imposing Eq.~\eqref{mb}
further entails appreciably reduced bottom-quark masses with 
correspondingly smaller $\Upsilon$-equilibrium numbers (see full 
line in Fig.~\ref{fig_open}).  
This, in turn, leads to $N_{\Upsilon}^{\text{eq}}$ 
being well below the initial hard production at both RHIC and LHC.
At the same time, the reaction rates are much
increased over the free gluo-dissociation scenario of the
previous section. 

\begin{figure}[!tbp]
  \includegraphics[width=0.9\linewidth,clip=]{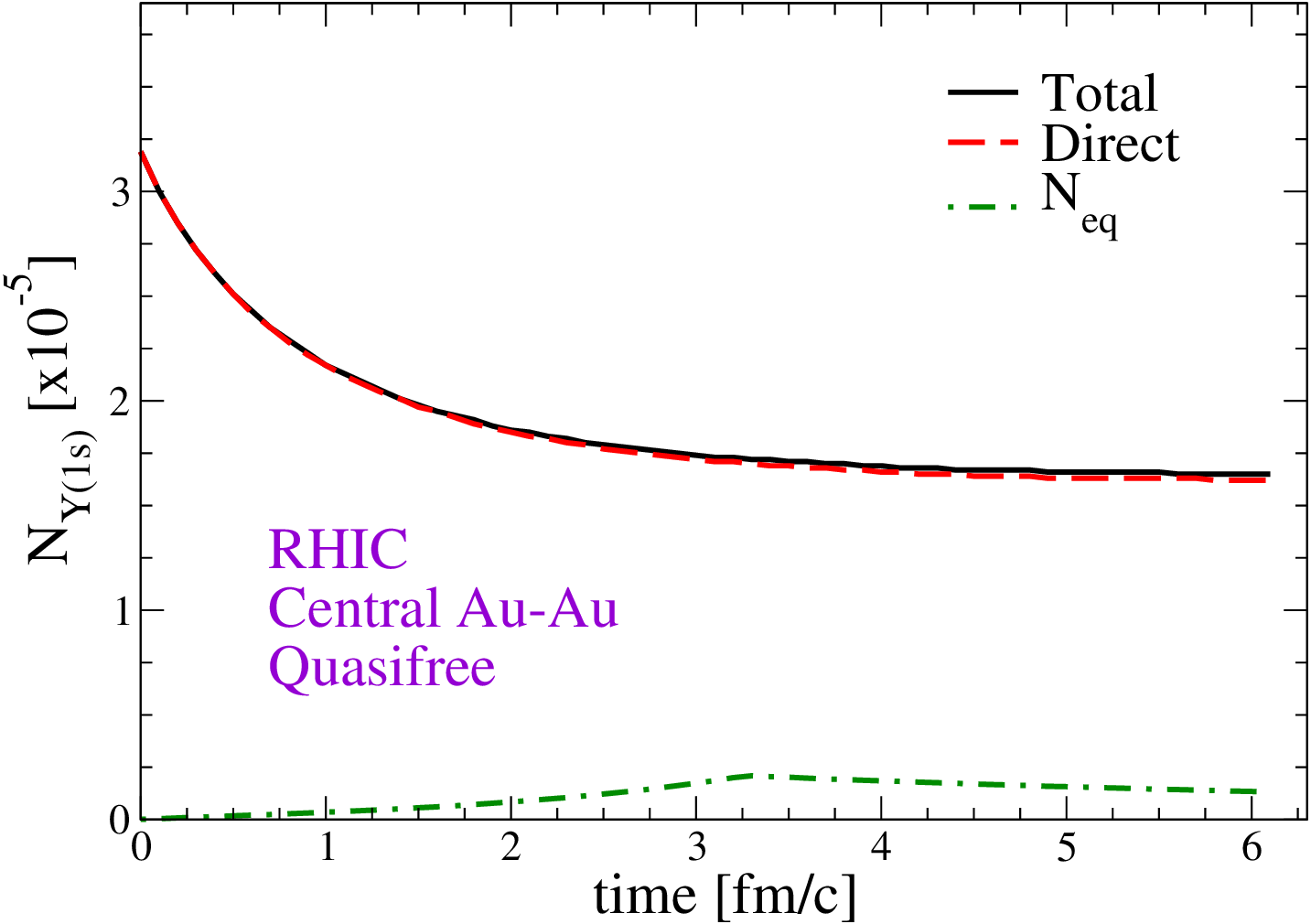}
  \caption{(Color online) $\Upsilon$(1s) abundances as a function of
    time for central ($b=1$~fm) Au-Au collisions at RHIC using the
    quasifree suppression mechanism with in-medium reduced binding
    energies (no feeddown from higher bottomonia included).  The solid
    line ($\Upsilon$ suppression and regeneration) is similar to the
    dashed line ($\Upsilon$ suppression only). This is a consequence of
    the small $\Upsilon$-equilibrium abundances (dash-dotted line).}
  \label{fig:gluo-evol-quasi-rhic}
\end{figure}
At RHIC, the $\Upsilon$-equilibrium number is so small that
secondary production is virtually absent, cf. dashed vs. solid line in
Fig.~\ref{fig:gluo-evol-quasi-rhic}.
However, the increased inelastic reaction rates lead to an appreciable 
suppression of nearly 50\% of the
initial number (after nuclear absorption). This rather dramatic effect 
is almost entirely due to the reduced $\Upsilon$ binding energy induced 
by a screening of the $b$-$\bar b$ potential in the QGP (the binding 
energy is already reduced to $\sim$550~MeV at $T_c$, rapidly 
dropping further before reaching zero at approximately 4.3~$T_c$). 
Thus, the magnitude of $\Upsilon$ suppression at RHIC appears
to be a rather sensitive measure for Debye screening
of the in-medium heavy-quark potential.  
\begin{figure}[!tbp]
  \includegraphics[width=0.9\linewidth,clip=]{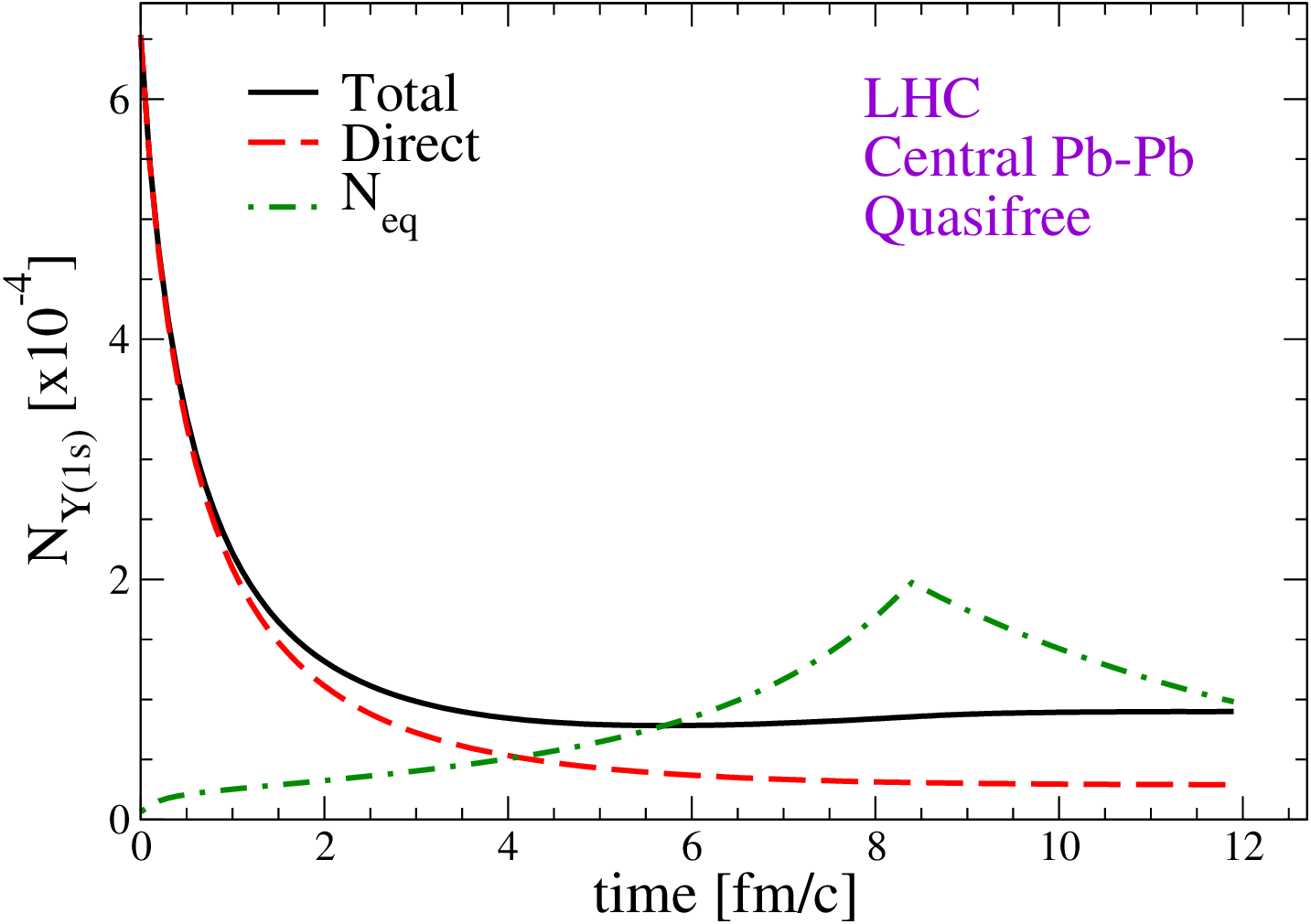}
  \caption{(Color online) The same as
    Fig.~\ref{fig:gluo-evol-quasi-rhic} but for central ($b=1$~fm) Pb-Pb
    collisions at LHC. While the total yield (solid line) builds up a
    large fraction of regenerated $\Upsilon$'s (due to noticable
    equilibrium abundances (dash-dotted line)), the prevalent feature is
    a net suppression by about a factor of 7.}
  \label{fig:gluo-evol-quasi-lhc}
\end{figure}
The effects become stronger at LHC, cf.~Fig.~\ref{fig:gluo-evol-quasi-lhc}.
Despite a significant increase of the final $\Upsilon$ number due to
regeneration in central collisions (solid vs. dashed line), the overall
effect is an almost seven-fold suppression of the
primordial yield (after nuclear absorption), which again is mostly
facilitated by an increased reaction rate originating from a reduced
binding energy, especially in the early phases.

\begin{figure}[!tbp]
  \includegraphics[width=0.9\linewidth,clip=]{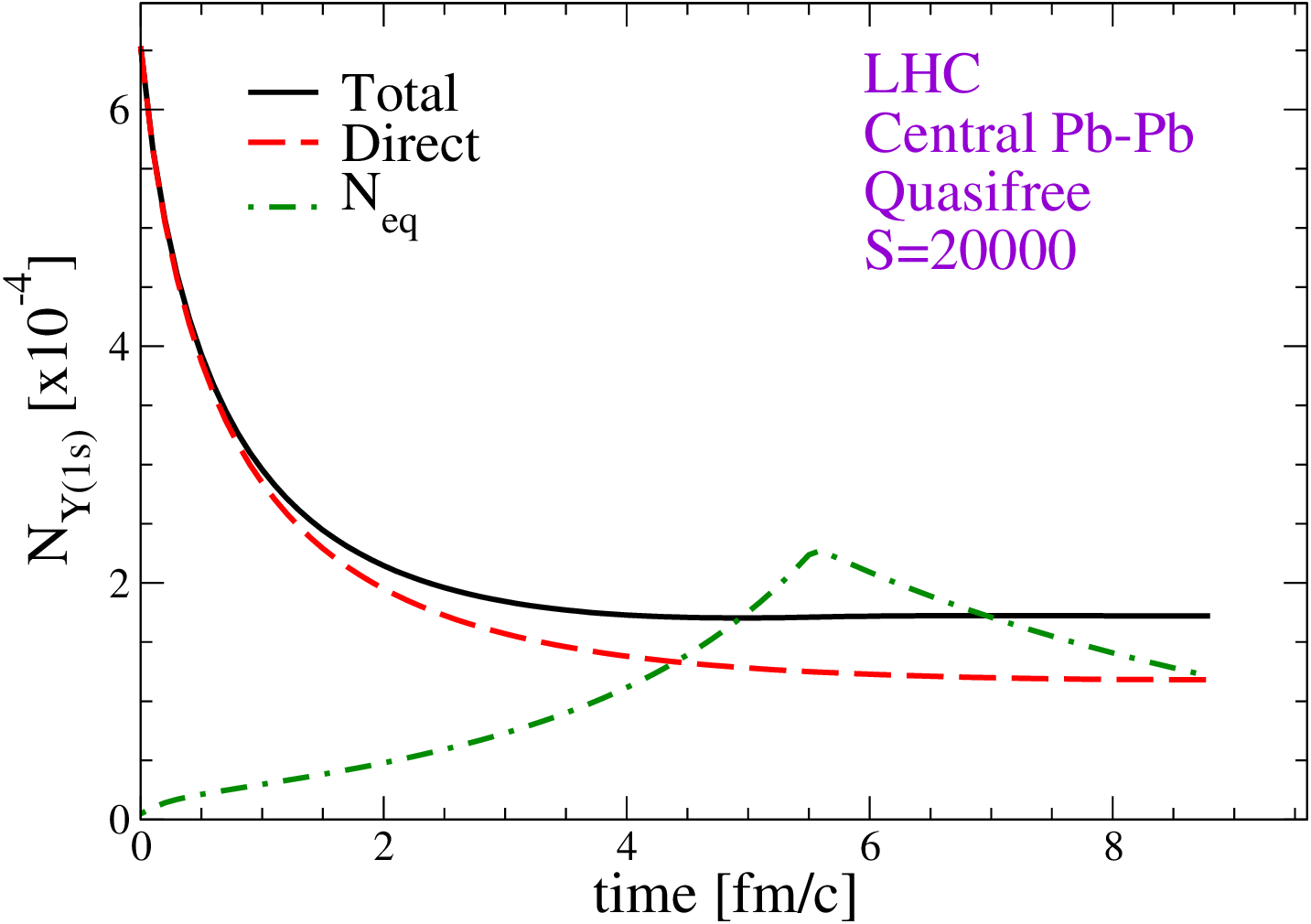}
  \caption{(Color online) Same as Fig.~\ref{fig:gluo-evol-quasi-lhc} but
    with a total entropy of the fireball reduced to half ($S$=20000),
    corresponding to a charged-particle multiplicity per unit rapidity
    of about $\sim$1600.}
  \label{fig:gluo-evol-quasi-lhc-20000}
\end{figure}
We furthermore check the sensitivity of our LHC calculation to the input
of the charged-particle multiplicity which is controlled by the total
entropy in the fireball, cf.~App.~\ref{app_fireball}. Reducing $S$ (and
therefore $\dd N_{\text{ch}}/\dd y$) by a factor of 2 and keeping
everything else the same amounts to smaller initial temperatures (by a
factor $\sim2^{1/3}$), shorter lifetimes of QGP (5.5~fm/c vs. 8.5~fm/c;
the duration of the mixed phase is changed very little) and smaller
volumes. The latter imply larger open-bottom densities at given $T$ and
thus an increase (by a factor of 2) of the open-bottom fugacity,
$\gamma_b$, which in turn increases the equilibrium limit of the
bottomonium number by that same factor. On the other hand, the smaller
temperatures and lifetimes of the evolution also inhibit the thermal
relaxation of bottom quarks, which reduces the correction factor ${\cal
  R}$ of Eq.~\eqref{relax}. The combination of these two effects leads
to an equilibrium number which both in magnitude and time-dependence is
quite similar to the evolution scenario with the larger $\dd
N_{\text{ch}}/\dd y \simeq 3200$ ($S$=40500). The main net effect is
thus the less efficient dissociation due to the smaller temperatures in
the evolution leading to a final suppression factor (relative to the
initial yield after nuclear absorption) of $\sim$1/4 for $\dd
N_{\text{ch}}/\dd y \simeq 1600$ as compared to $\sim$1/7 for $\dd
N_{\text{ch}}/\dd y \simeq 3200$. Furthermore, the final yield is now
dominated by the suppressed direct contribution.

We also note that at both RHIC and LHC most of the suppression
occurs in the first 2-3~fm/c, underlining that $\Upsilon$ production 
is an observable which is sensitive to early (QGP) phases of 
central heavy-ion reactions.  

\subsubsection{Comparison to Charmonium}
\label{subsubsec_charm}

It is worthwhile to confront the bottomonium results with the ones for
charmonium of our previous analysis along similar
lines~\cite{GR01,GRB04}: when going from SPS to RHIC energies, we have
predicted a transition from a charmonium suppression-dominated regime at
SPS to mostly regeneration at RHIC, with an approximately constant net
suppression relative to initial production. The situation for
bottomonium found here is quite different: in scenarios with color
screening we expect the dominant mechanism to be $\Upsilon$ suppression
both at RHIC and LHC, which increases with collision energy. This
suggests the interesting possibility of observing an absence of
suppression (or even enhancement over initial production~\cite{An03})
for charmonium at LHC~\cite{An03} together with substantial suppression
of bottomonium which would provide a rather unambiguous signature of
charmonium regeneration in particular, and QGP formation (including
screening) in general.  A similar feature (albeit less pronounced) might
also be present at RHIC if color-screening and associated reduction in
$\Upsilon$ binding energies become operative below temperatures of $\sim
2T_c$: in central Au-Au a 50\% suppression of (the much more tightly
bound) $\Upsilon$ (after nuclear absorption) should be compared to a
30\% net suppression of $J/\psi$'s after inclusion of the (prevalent)
regeneration.

We emphasize again that the magnitude of the net $\Upsilon$ suppression
is quite sensitive to the strength of color screening (affecting the
bottomonium binding energies, and, in turn, the $b$-quark mass according
to $m_b=(m_Y+\varepsilon_B)/2$): at both RHIC and LHC the scenario
without screening (Figs.~\ref{fig:gluo-evol-std-rhic} and
\ref{fig:gluo-evol-std-lhc}) leads to little net suppression (contrary
to Figs.~\ref{fig:gluo-evol-quasi-rhic} and
\ref{fig:gluo-evol-quasi-lhc} where screening is included). At RHIC,
this is due to the larger binding energies leading to a factor of
$\sim$10 decrease in the reaction rates around $T=300$~MeV, while at LHC
it is due to larger $b$-quark masses implying larger $Y$ equilibrium
abundances which facilitate the regeneration in the early phases.  On
the other hand, uncertainties in the implementation of the screening
mass (\eg, using $\mu_D=\sqrt{1+N_f/6} \, gT$ with $N_f$=3 rather
than the default $\mu_D=gT$) are small (less than 15\% for the final
abundances).  We recall that all considerations so far concern the
exclusive $\Upsilon$ yields in central $A$-$A$ collisions.

\subsection{Excited Bottomonia and Feeddown}
\label{subsec_feed}
A complete description of inclusive $\Upsilon$ production requires
to account for feeddown contributions from (late) decays of
higher bottomonium states. Thus, one also needs to determine the 
(in-medium) production systematics of these states. In addition, it is 
expected that some excited bottomonia may be measured directly, 
providing further model tests.
According to Ref.~\cite{YELL03} and previous experimental studies
of bottomonium systems at Fermilab~\cite{Affolder:1999wm}, the final
$\Upsilon$ yield observed in $p\bar{p}$ collisions at
$\sqrt{s_{NN}}=39$~GeV is composed of up to 50\% from feeddown, 
see Tab.~\ref{tab:feedown}. 
\begin{table}[!tbp]
  \centering
  \begin{tabular}{| l | c |}
\hline
prompt $\Upsilon(1s)$ & $\sim 51\%$, \\ \hline
$\Upsilon(1s)\ \mathrm{from}\ \chi_b(1P)\ \mathrm{decays}$ & $\sim
27\%$,\\ 
$\Upsilon(1s)\ \mathrm{from}\ \chi_b(2P)\ \mathrm{decays}$ & 
$\sim 10\%$,\\
$\Upsilon(1s)\ \mathrm{from}\ \Upsilon(2s)\ \mathrm{decays}$ & $\sim
 11\%$,\\  
$\Upsilon(1s)\ \mathrm{from}\ \Upsilon(3s)\ \mathrm{decays}$ & $\sim
1\%$.\\ \hline
  \end{tabular}
  \caption{Decomposition of the inclusive $\Upsilon$
    yield into prompt and feeddown contributions from excited bottomonia
    in $p\bar{p}$ collisions at $\sqrt{s_{NN}}=39$~GeV~\cite{Affolder:1999wm}.}
  \label{tab:feedown}
\end{table}
For simplicity, we will in the following neglect $\Upsilon$(3s) states 
and will not distinguish between $\chi_b$ states. We will also assume 
that the composition of the primordial
$\Upsilon$ yield is energy- and system-size independent.

\begin{figure}[!tbp]
  \centering
  \includegraphics[width=0.9\linewidth,clip=]{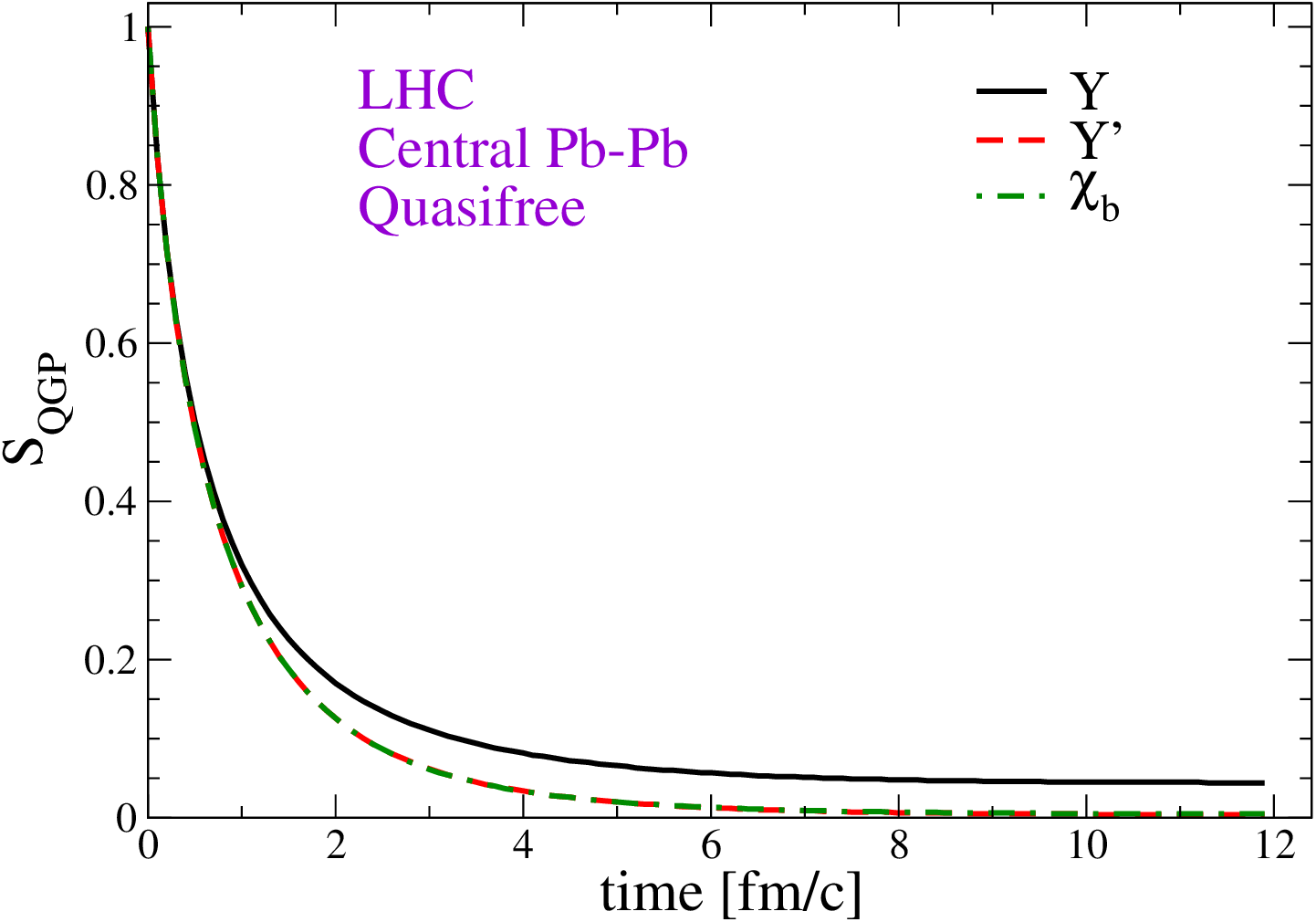}
  \caption{(Color online) QGP suppression of bottomonia in central
    ($b=1$~fm) Pb-Pb collisions at LHC, using the quasifree destruction
    process. Solid line: $\Upsilon$, dashed line: $\Upsilon'$,
    dash-dotted line: $\chi_b$.}
  \label{fig:suppression-1s-2s-1p}
\end{figure}
The dissociation rates for $\Upsilon'$ and $\chi_b$ are evaluated within
the quasifree formalism including in-medium binding energies following
the screening estimates of Ref.~\cite{KMS88} (amounting to
$\varepsilon_B\simeq$150-200~MeV at $T_c$, reaching zero at about
1.8~$T_c$), cf. Figs.~\ref{fig:Ediss} and \ref{fig:rates-quasifree}
(recall from Sec.~\ref{subsec_life} that the gluo-dissociation approach
cannot be applied to the small binding energies at hand and therefore
does not allow for a consistent treatment of feeddown effects).  For
illustration purposes, we display in Fig.~\ref{fig:suppression-1s-2s-1p}
the pertinent suppression factors, Eq.~\eqref{re.4}, for central Pb-Pb
collisions at LHC (neglecting the gain term in the rate equation
\eqref{rate-eq}). The $\Upsilon'$ and the $\chi_b$ exhibit similar
suppression (dashed and dash-dotted ($\chi_b$) line, respectively),
almost a factor of 10 stronger than for $\Upsilon$ mesons (solid line),
which is, of course, due to their smaller binding energies.

The much stronger suppression of excited bottomonia (as well as
their smaller equilibrium limits) compared to direct $\Upsilon$'s 
has important consequences for the inclusive $\Upsilon$ yields at  
both RHIC and LHC which will be shown in the following section. For
central $A$-$A$ collisions, the suppression in inclusive $\Upsilon$
production is about 30-40\% more pronounced than for
exclusive yields as displayed in Sec.~\ref{sec_kin}.

\section{Centrality Dependence}
\label{sec_centr}

The results presented in the previous sections were calculated for
central collisions ($b=1$~fm).  In this section, we extend our results
to different centrality classes by solving the rate Eq.~\eqref{rate-eq}
for Au-Au (Pb-Pb) collisions for various impact parameters, $b$, (as was
done before for charmonia in Ref.~\cite{GRB04}), and then plotting the
yields at the end of each time evolution as a function of the number of
nucleon participants, $N_{\rm part}(b)$.

To solve the rate equations, we focus on the inelastic reaction rates
obtained from the quasifree suppression mechanism for $b$-quarks in the
QGP with in-medium mass $m_b(T)$ (corresponding to the lower set of
curves in Fig.~\ref{fig:rates-quasifree}) and include thermal
off-equilibrium corrections for $N_\Upsilon^{\rm eq}(t)$ (according to
eq.~\eqref{relax}). The inclusive $\Upsilon$ yields account for the
feeddown from $\Upsilon'$ and $\chi_b$ for which we solve the same rate
Eq.~\eqref{rate-eq} with pertinent reaction rates.  The primordial
numbers of $b\bar{b}$ pairs and bottomonia (the latter subjected to an
identical but centrality-dependent nuclear absorption) are obtained by
scaling our input numbers of Sec.~\ref{sec_hard} according to the number
of binary collisions, $N_{\text{coll}}(b)$ (characteristic for hard
processes).

The centrality dependence of the fireball evolution
(App.~\ref{app_fireball}) is constructed assuming the total entropy at
given impact parameter to be proportional to $N_{\text{part}}(b)$,
$S_{\text{tot}}(b)\propto N_{\text{part}}(b)$ (characteristic for soft
processes, which for RHIC data is approximately satisfied within 20\% or
so; for LHC the uncertainty is larger due to competing effects
anticipated from gluon saturation and an increasing importance of a
$N_{\text{coll}}$-component~\cite{KLN05}). The initial transverse radius
($r_0$) is estimated in cylindrical symmetry from the nuclear overlap
function with constant formation time, tranverse acceleration and
longitudinal velocity, cf.~Eq.~\eqref{fb.1} (for the purpose of
extracting the temperature and volume evolution, effects of finite
ellipticities in noncentral collisions are not significant).

In the following, we first study the centrality dependence
of the $\Upsilon$ yield at RHIC, then at LHC, and finally the 
$\Upsilon'/\Upsilon$ ratio.

\subsection{RHIC}
\label{sec:centr-depend-rhic}
\begin{figure}[!tb]
  \centering
  \includegraphics[width=0.9\linewidth,clip=]{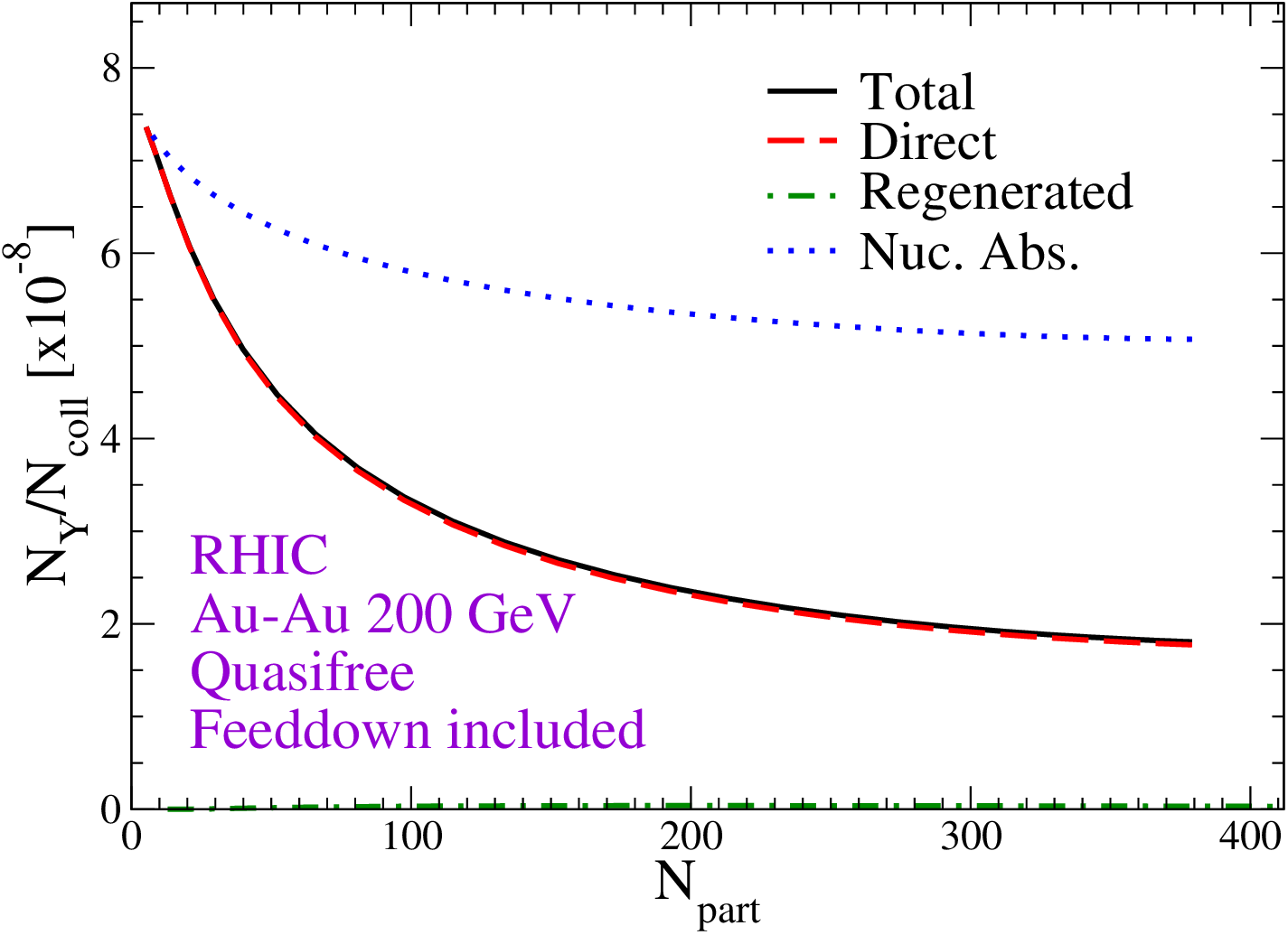}
  \caption{(Color online) Centrality dependence of
    $N_{\Upsilon}/N_{\text{coll}}$ at RHIC (Au-Au collisions at
    $\sqrt{s_{NN}}=200$~GeV) using the quasifree suppression mechanism
    including color-screening. Dashed line: suppressed primordial
    $\Upsilon$. Dash-dotted line: regenerated $\Upsilon$.  Solid line:
    total $\Upsilon$.}
  \label{fig:centrality-rhic}
\end{figure}
Fig.~\ref{fig:centrality-rhic} shows our results for $\Upsilon$
production in $\sqrt{s_{NN}}=200$~GeV Au-Au as a function of 
participant number, normalized to the number of binary collisions. As
expected from the dependencies in Fig.~\ref{fig:gluo-evol-quasi-rhic}, 
the $\Upsilon$ yield (solid line) exhibits a substantial increase in 
suppression going from peripheral to central collisions, with 
negligible regeneration (dash-dotted line), \ie, the final $\Upsilon$ 
yield is exclusively due to primordial production 
subject to suppression in the QGP (dashed line). This is
primarily a consequence of the small number of $b\bar{b}$ pairs 
created in hard nucleon-nucleon scatterings, implying an $\Upsilon$ 
equilibrium number which is well below the actual
number at any time (more so toward peripheral collisions).
The prevalent effect is thus a significant suppression of the inclusive
$\Upsilon$ yield with centrality, originating from (i) suppression of direct 
$\Upsilon$'s due to (substantial) color screening, and (ii) suppression of
(less bound) excited bottomonia with subsequent feeddown. 
Overall, our findings for $\Upsilon$ production at RHIC are rather 
reminiscent of the $J/\Psi$ suppression pattern at SPS as calculated
in a similar approach~\cite{GRB04} 
(in line with NA50 data~\cite{Alessandro:2004ap}).

\subsection{LHC}
\label{sec:centr-depend-lhc}
\begin{figure}[!tb]
  \centering
  \includegraphics[width=0.9\linewidth,clip=]{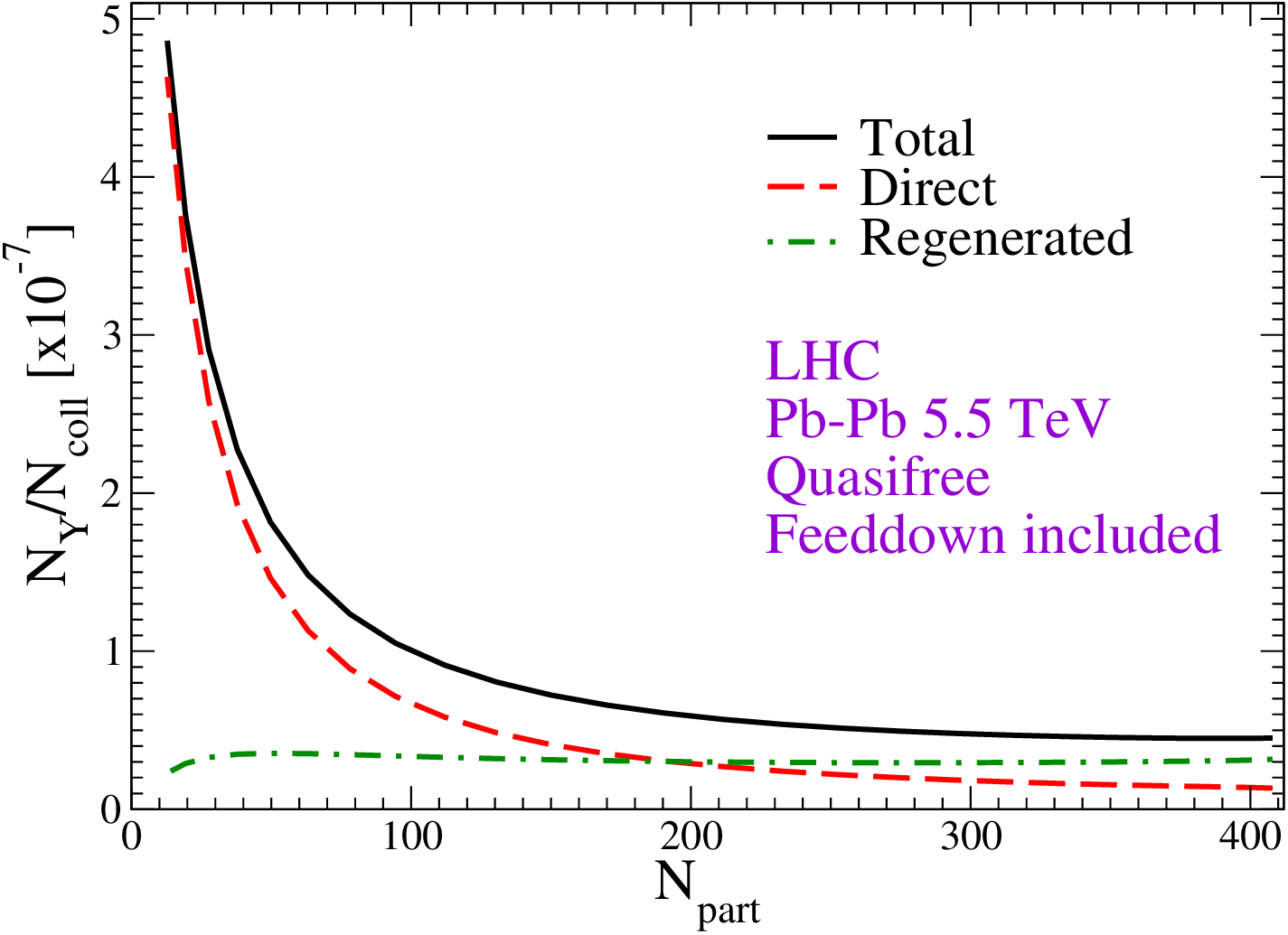}
  \caption{(Color online) Same as Fig.~\ref{fig:centrality-rhic}, but
    for Pb-Pb collisions at LHC ($\sqrt{s_{NN}}=5.5$~TeV) with $\dd
    N_{\text{ch}}/\dd y|_{\rm central}\simeq 3200$.}
  \label{fig:centrality-lhc}
\end{figure}

In Fig.~\ref{fig:centrality-lhc} we present our predictions for the
centrality dependence of the $\Upsilon$ yield at LHC within the same
approach as for RHIC. The longer lifetime of the QGP phase and,
in particular, the higher temperatures reached at LHC render the
suppression of primordial $\Upsilon$  (dashed line) even more pronounced. 
Regeneration (dash-dotted line), however, is no longer negligible, 
accounting for up to 2/3 of the final yield in central collisions.
\begin{figure}[!tb]
  \centering
  \includegraphics[width=0.9\linewidth,clip=]{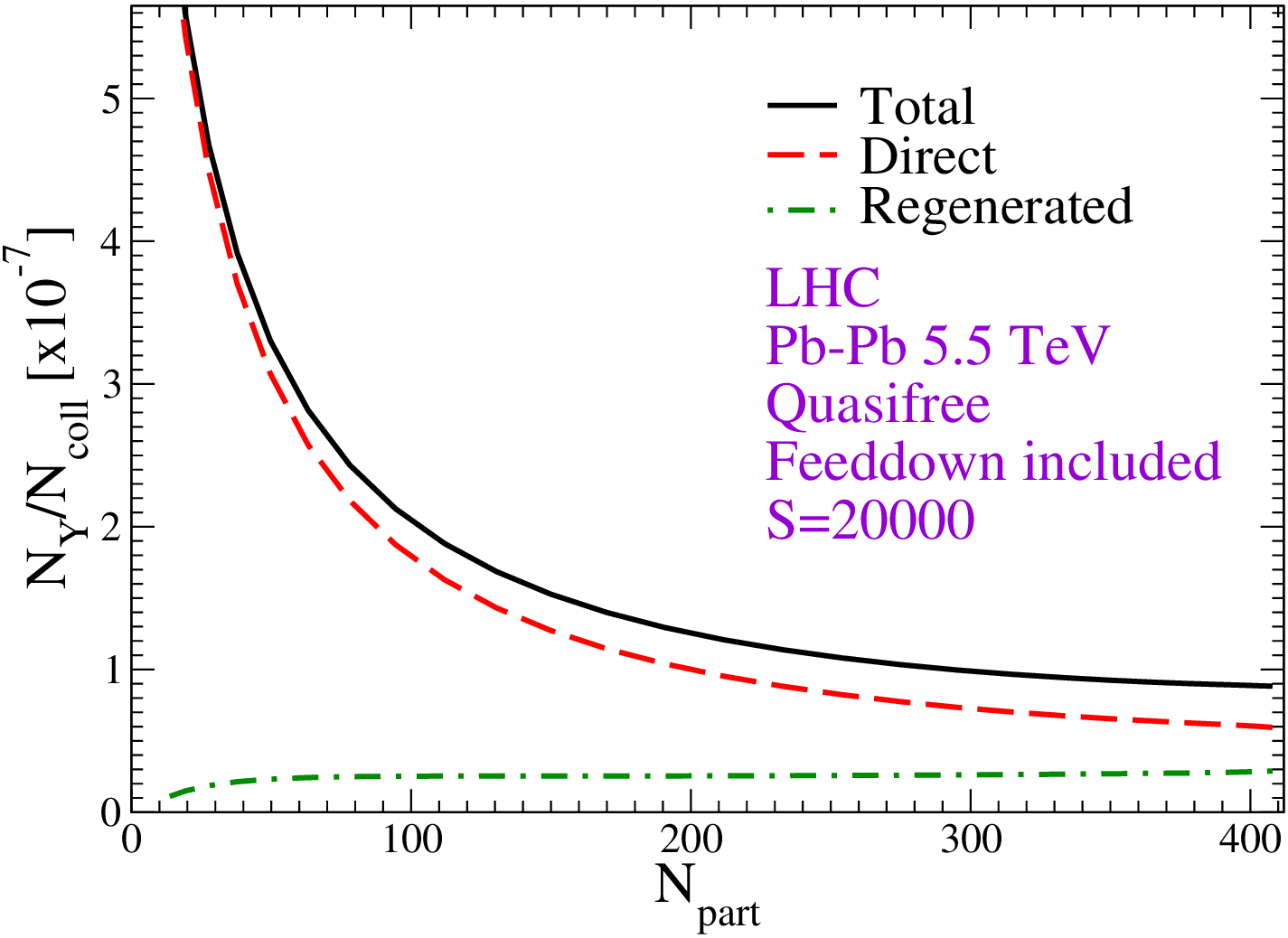}
  \caption{(Color online) Same as Fig.~\ref{fig:centrality-lhc}, but
    with half the total entropy in the fireball at each impact
    parameter, corresponding to $\dd N_{\text{ch}}/\dd y \simeq 1600$ in
    central collisions.}
  \label{fig:centrality-lhcS20000}
\end{figure}
This should leave further traces in $p_t$-dependencies, such as a
softening of the spectra and a sizable $v_2$.  Nevertheless, the
situation at LHC is qualitatively similar to the one at RHIC as far as
the centrality dependence is concerned, \ie, the prevailing
manifestation of QGP formation is $\Upsilon$ suppression at
\textit{both} RHIC and LHC.

We also illustrate again the impact of a smaller charged-particle
multiplicity in Pb-Pb at LHC, leaving all other quantities unchanged, see
Fig.~\ref{fig:centrality-lhcS20000}.  As found for the time dependence in
central collisions (Sec.~\ref{subsubsec_quasifree}), the smaller initial
temperatures and QGP lifetimes lead to a factor of $\sim$2 less
suppression. Also note that the absolute number of regenerated $\Upsilon$'s
hardly changes, implying that the final yield is now dominated by the
primordial component.

\subsection{$\Upsilon'/\Upsilon$ ratio}
\label{sec:up-up-ratio}
\begin{figure}[!tbp]
  \centering
  \includegraphics[width=0.9\linewidth,clip=]{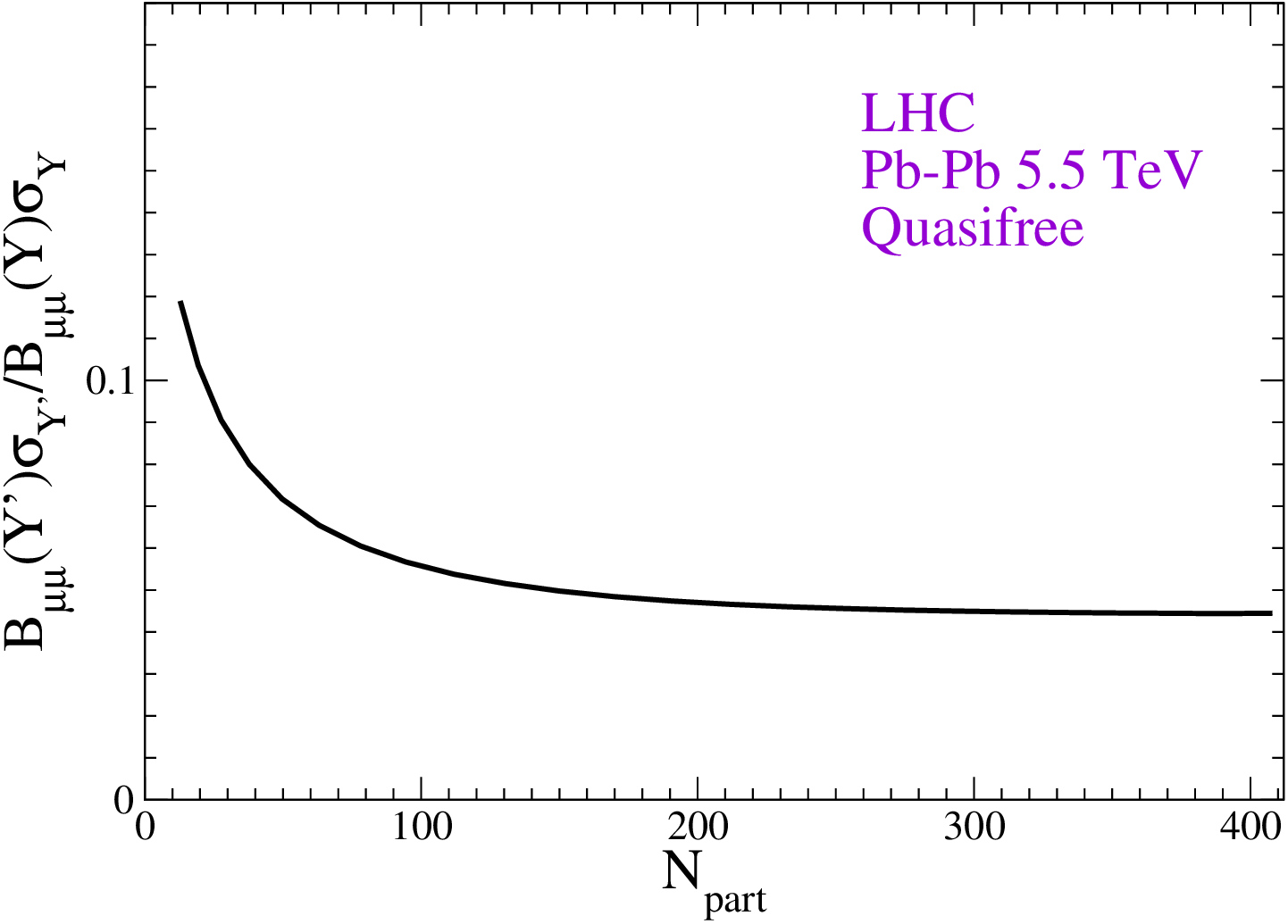}
  \caption{(Color online) Centrality dependence of
    $N_{\Upsilon'}/N_{\Upsilon}$ (supplemented by the dilepton branching
    ratios of $B_{\mu \mu}(\Upsilon')$$\sim$1.3\% and $B_{\mu
      \mu}(\Upsilon)$$\sim$2.5\%, respectively; feeddown on
    $\Upsilon'$ states is not included) at LHC for Pb-Pb collisions at
    $\sqrt{s_{NN}}=5.5$~TeV.}
  \label{fig:centrality-2s-1s}
\end{figure}

We finally extract the $\Upsilon'/\Upsilon$ ratio at LHC which was
already implicit in the centrality dependence of the inclusive
$\Upsilon$ yield shown in the previous section. The smaller $\Upsilon'$
binding energy obviously facilitates its suppression, and its larger
mass additionally inhibits regeneration (due to smaller equilibrium
abundances).  Consequently, we find a decreasing $\Upsilon'/\Upsilon$
ratio, by about a factor of 2 (3) when going from peripheral ($p$-$p$)
to central collisions, cf.~Fig.~\ref{fig:centrality-2s-1s} (feeddown to
the $\Upsilon'$ has been neglected; note that the limit for $p$-$p$
amounts to 33\%, or 17\% upon inclusion of the dilepton decay branching
ratios).  We note that even for central collisions, our calculations
amount to a ratio of $\Upsilon'/\Upsilon\sim$9\% which lies well above
its thermal-equilibrium value of $\sim$5\% (at $T=180$~MeV) that one
would expect if bottomonium states were entirely created via statistical
hadronization~\cite{Becattini:2005hb}. The reason is that the number of
$\Upsilon'$ reaches its terminal value at the beginning of the mixed
phase where the pertinent reaction rate, $\Gamma_{\Upsilon'}$, becomes
small, while the equilibrium limit drops by a factor of two toward the
end of the mixed phase (as for the $\Upsilon$ in
Fig.~\ref{fig:gluo-evol-quasi-lhc}). On the other hand (see also
Fig.~\ref{fig:gluo-evol-quasi-lhc}), the terminal number of $\Upsilon$
essentially happens to coincide with the equilibrium value at the end of
the mixed phase (even though it is frozen much earlier). Therefore the
final $\Upsilon'$/$\Upsilon$ ratio is roughly a factor two larger than
the equilibrium value at the end of the mixed phase, which is the result
displayed in Fig.~\ref{fig:centrality-2s-1s} for central collisions.

\section{Conclusions}
\label{sec_concl}
At heavy-ion colliders (RHIC and LHC) hadrons containing bottom quarks
are expected to become an important new probe of the produced (novel)
forms of strongly interacting matter. In this article we have presented
a kinetic theory framework to assess the evolution of bottomonia in the
QGP phases of high-energy heavy-ion reactions, neglecting any
reinteractions in hadronic matter (which is presumably a fair
approximation for $\Upsilon$, $\chi_b$ and $\Upsilon'$ states).  The
underlying rate equation requires several inputs: (1) initial
bottomonium abundances (which we estimated from collision-scaled $p$-$p$
data), (2) space-time evolution of temperature (for which we employed an
expanding fireball model), (3) inelastic reaction rates of bottomonia,
and (4) bottomonium equilibrium abundances.

For the inelastic reaction rates we have evaluated two mechanisms for
dissociation due to thermal light quarks and gluons in the QGP: (i)
gluo-absorption (the analogue of photo-dissociation) and (ii)
``quasifree'' scattering of both quarks and gluons off individual $b$
and $\bar b$-quarks in the bound state.  Whereas processes of type (i)
are appropriate (and expected to be dominant) for deeply bound
quarkonia, processes (ii) allow to treat the case of small binding
energies (where gluo-dissociation leads to anomalously small reaction
rates).  Consequently, when introducing effects of color-screening, the
associated (temperature-dependent) reduced binding energies mandate the
use of the quasifree processes, while for vacuum binding energies
gluo-dissociation is employed.

The equilibrium limits of bottomonia, which govern the gain term 
(regeneration) in the rate equation, follow the trends previously 
established for charmonia: for a fixed number of $b\bar b$ pairs in 
the system (at given centrality), larger masses (and smaller 
degeneracies) of open-bottom states imply larger $Y$-equilibrium 
limits (rendering bottomonium formation more favorable).  However, 
in most cases considered (in particular when implementing 
medium effects with relatively small $b$-quark masses) the 
equilibrium limit was significantly below the initial value from hard
production (after nuclear absorption), especially after applying a
schematic correction for incomplete $b$-quark thermalization.

Our main results may therefore be summarized as follows: (I)
color-screening (\ie, in-medium reduced binding energies) leads to a
substantial increase of the inelastic bottomonium reaction rates; (II)
at RHIC, this manifests itself as a 50\% suppression effect (factor
$\sim$2) in the QGP phase (with virtually no regeneration), whereas for
vacuum $\Upsilon$ states almost no (thermal) suppression occurs (in
which case suppression in inclusive yields is due to feeddown from
excited states); (III) at LHC, assuming $\dd N_{\text{ch}}/\dd y \simeq
3200$ in central Pb-Pb, the net QGP suppression becomes substantially
stronger (factor $\sim$7 (1.4) for in-medium (vacuum) binding energies),
with the final yield containing a 75\% (30\%) component of regenerated
$\Upsilon$'s. If the charged-particle multiplicity turns out to be
smaller (\eg, $\dd N_{\text{ch}}/\dd y\simeq 1600$), the reduced QGP initial
temperature and lifetime will lead to weaker net suppression (\eg,
factor $\sim$2 less).

Our general finding is thus that in heavy-ion reactions at \textit{both}
RHIC and LHC, $\Upsilon$ suppression is the prevalent effect as a
function of collision centrality, and there is large sensitivity to
color-screening in the QGP.  This is quite different from our previous
studies of charmonia where $J/\Psi$ regeneration is expected to be the
dominant mechanism for RHIC energies and above.
 
Concerning future improvements, it is desirable to provide a more 
accurate treatment of in-medium bottomonium properties, implementing
both screening and dissociation processes, as well as non-thermal
$b$-quark distributions, simultaneously in a more microscopic approach.  

\vspace{0.3cm}

\acknowledgments 
This work was supported in part by the U.S. National Science Foundation
through a CAREER award (RR) under Grant No. PHY-0449489 and 
through a REU program (SL) under Grant No. PHY-0354098.
One of us (HvH) thanks the Alexander von Humboldt foundation for
partial support as a Feodor Lynen fellow.  
One of us (LG) was supported by the Director, Office of Energy
Research, Office of High Energy and Nuclear Physics, Divisions of
Nuclear Physics, of the U.S. Department of Energy under Contract No.
DE-AC02-05CH11231.

\begin{appendix}
\section{Brief Review of Lattice QCD Results}
\label{app_lqcd}

Finite-temperature lQCD calculations exhibit a substantial reduction of
the heavy-quark free energy~\cite{Karsch:2000kv,Petreczky:2004pz},
\begin{equation}
F_{Q\bar{Q}}(r,T) = U_{Q\bar{Q}}(r,T) - T \, S_{Q\bar{Q}}(r,T) \ ,
\end{equation}
with increasing temperature. When implementing the free energy into a
Schr\"odinger equation, $J/\psi$ ($\psi'$) mesons have been found to
dissolve slightly above (below) $T_c$~\cite{Digal:2001ue}. More
recently, it has been argued that the internal energy, $U_{Q\bar Q}$, is
the more appropriate quantity to be identified with a potential.
Corresponding solutions to the Schr\"odinger equation entail
significantly larger dissociation temperatures of the $J/\psi$ of around
$T_{\text{diss}}$$\simeq
2T_c$~\cite{Wong:2004zr,SZ04,MR05,Alberico:2005xw}.  This led, in fact,
to better consistency with (quenched) lQCD results for charmonium
spectral functions, in which the peak structures associated with the
lowest-lying resonances ($J/\psi$ and $\eta_c$) disappear slightly below
2$T_c$~\cite{Umeda:2002vr,Datta:2003ww,Asakawa:2003re}.  With the mass
(peak position) of the spectral functions approximately staying
constant~\cite{Umeda05}, the key property of the charmonium spectral
functions is thus a change in their widths, \ie, inelastic reaction
rates.  This is further corroborated by lQCD studies which find the
$J/\psi$ width to increase across the phase transition, from very small
below to about 200~MeV just above $T_c$~\cite{Umeda05}.  Such a behavior
is quite consistent~\cite{RG03} with both hadronic model calculations
below $T_c$, which typically result in small inelastic reaction rates of
$\lsim$0.1~fm$^{-1}$ (mostly due to $\pi, \rho +J/\psi$
breakup)~\cite{Duraes:2002ux}, and parton-induced destruction above
$T_c$, typically of the order of 1~fm$^{-1}$.  Also note that in the
presence of color screening, which induces a reduction in the $c$-$\bar
c$ binding energy ($\epsilon_B>0$), a constant charmonium mass implies a
decrease in the open-charm threshold, according to the relation
\begin{equation}
m_\Psi = 2 m_c - \varepsilon_B \ .
\label{mc}
\end{equation}
A decrease in the open-charm threshold with $T$ has indeed been found
in the lQCD free energies (or potentials), at least above
$T_c$~\cite{Kaczmarek:2005gi}, and
complies naturally with an increase in the bound state width, due
to larger phase space in the break-up reactions, \eg,
$g+J/\psi \to c + \bar c$.

Effects of the hadronic phase are neglected in this article; first lQCD
results~\cite{Petrov:2005sd} (as well as potential models based on
lQCD~\cite{Wong:2004zr}) indeed indicate much higher dissociation
temperatures of the $\Upsilon$ than for $J/\psi$, around
$T_{\text{diss}}^{\Upsilon} \simeq 4T_c$.

\section{Thermal Fireball Parameterization}
\label{app_fireball}

The starting point for the thermal fireball model is a cylindrical
ansatz for the (proper) time dependence of the (isotropic) three-volume
according to
\begin{equation}
\label{fb.1}
V_{\text{FB}}(t)=\left (z_0+v_z t \right) \pi \left
  (r_0+\frac{a_{\perp}}{2} t^2 \right)^2 \ . 
\end{equation}
The longitudinal expansion is characterized by (i) an expansion velocity
of $v_z=1.4c$ representing a rapidity range of $\Delta y \simeq 1.8$
typical for a thermal fireball width, and (ii) an initial length
$z_0=0.6 {\ } (0.125)$~fm for RHIC (LHC) which carries the meaning of a
formation (thermalization) time, $\tau_0=z_0/\Delta y$ =0.33 (0.07)~fm/c
at RHIC (LHC). The initial transverse size is determined by the
centrality of the collision, \eg, $r_0=6.4$~$(6.6)$~fm for $b=1$~fm Au-Au
(Pb-Pb) collisions at RHIC (LHC).

If one further assumes the total entropy, $S$, of the system to
be conserved (\ie, isentropic expansion), the temperature evolution 
of the fireball matter can be inferred by equating  
the entropy density to the one of a thermal medium. For a 
non-interacting QGP one has 
\begin{equation}
\begin{split}
\label{entro}
s(t) &=\frac{S}{V_{\text{FB}}(t)} \\ & 
\equiv \mp \sum_{i} d_i \int
\frac{\dd^3 k}{(2 \pi)^3} \big \{1 \pm f_i(\omega_i) \ln[f_i(\omega_i)]
\\ & \quad \quad +[1 \mp f_i(\omega_i)] \ln[1\mp f_i(\omega_i)] \big
\} \ ,  
\end{split} 
\end{equation}
where $f_{i}$ denote the Fermi-Dirac ($i$=$q$, $\bar q$, upper signs)
or Bose-Einstein ($i$=$g$, lower signs) distribution functions of
(massive) quarks or gluons with on-shell energies
$\omega_i=\sqrt{k^2+m_i^2}$ ($m_{i}$: thermal mass), and $d_i$ stand for
the pertinent spin-color-flavor degeneracies.  We will consider two
scenarios for the QGP equation of state: (i) massless partons with an
effective number of quark flavors $N_f=2.5$; in this case,
$s=(d_g+10.5N_f) 4\pi^2/90 T^3$, and the temperature can be obtained
explicitly as $T(t)=\text{const}\times(S/V_{FB}(t))^{1/3}$; (ii) a
three-flavor QGP with thermal parton masses,
\begin{equation}
\label{masses}
m_{u,d}^2 = \frac{g^2 T^2}{6}, \quad m_s^2=m_0^2+\frac{g^2 T^2}{6}, 
\quad m_g^2=\frac{3}{4}g^2 T^2,
\end{equation}
where $m_0$=150~MeV is taken as a bare strange-quark mass; in this
scenario, Eq.~(\ref{entro}) provides an implicit relation for $T(t)$
which is solved numerically. 
However,  we observe no significant
differences for the temperature profiles when calculating the
entropy density for a QGP with massless degrees of freedom (and 
$N_f=2.5$) or massive partons according to Eq.~(\ref{masses}).

If the entropy density, $s(t)$, drops below the critical value of the 
QGP phase, $s_c^{\text{QGP}}\equiv s^{\text{QGP}}(T_c=180~{\rm MeV})$,  
we perform the standard mixed-phase construction,
\begin{equation}
\label{mix}
\frac{S}{V_{\text{FB}}(t)}=f s_c^{\text{HG}}+(1-f) s_c^{\text{QGP}}
\end{equation}  
with volume fractions $f$ and 1$-$$f$ of matter in the hadron gas (HG)
and QGP, respectively. The entropy density of the HG at $T_c$,
 $s^{\text{HG}}(T_c)\simeq 6.1$~fm$^{-3}$, is computed from a HG 
including all resonances up to masses of about 2~GeV, with a
baryonic chemical potential $\mu_B\simeq$~30~(2)~MeV corresponding
to (expected) chemical freezeout at RHIC (LHC). The
charged-particle multiplicity is then determined by
fixing the total entropy of the system~\cite{Rap01}, 
\eg, $S$=10000 (40500) for
central collisions at RHIC (LHC) translating into 
$\dd N_{\text{ch}}/\dd y\simeq$~800 (3200; we have evaluated a LHC scenario with
$S$=20000, \ie, $\dd N_{\text{ch}}/\dd y\simeq$~1600, which is closer
to extrapolations based on present RHIC data~\cite{An03,Back:2004je}). 
As mentioned above, we neglect the hadronic evolution of the system.   

\section{Equilibrium Abundances}
\label{app_eq-abund}
Following our assumption of a fixed number of $b$-$\bar b$ pairs
(at given centrality) in the fireball, the $Y$-equilibrium number not only
depends on temperature and volume, but also on a bottom-quark fugacity
$\gamma_b\equiv\gamma_{\bar b}$ via
\begin{equation}
\label{Neq}
N_{Y}^{\text{eq}}(T) = d_Y \, \gamma_b(T)^2 \, V_{\text{FB}}(t) 
\int \frac{\dd^3 p}{(2 \pi)^3} \, \exp \left (-\frac{E_Y(p)}{T} \right) \, 
\end{equation}
($E_Y(p)=\sqrt{m_{Y}^2+p^2}$, $d_Y$: spin degeneracy).
The temperature dependence of the fugacity is determined by requiring   
that at each moment the total number of $b$- ($\bar b$-) quarks 
in open and hidden bottom states matches the primordially produced
abundance~\cite{Goren02}, 
\begin{equation}
\label{match}
N_{b\bar{b}}=\frac{1}{2} \gamma_b N_{\text{op}} \frac{I_1(\gamma_b
  N_{\text{op}})}{I_0(\gamma_b N_{\text{op}})} + \gamma_b^2
  N_{\text{hid}}\ ,
\end{equation}
where $I_j$ are modified Bessel functions. The numbers
of hadronic open and hidden bottom states are given by
\begin{eqnarray}
  N_{\text{op}}^B &=& V_{\text{FB}} \sum_{x} n_x \, , \ x \in
  \{B, \bar B, B^*, \bar B^*, \Lambda_b, \ldots \} ,
\label{NopH}
\\
  N_{\text{hid}} &=& V_{\text{FB}} \sum_{y} n_y \, , \ y \in
  \{\Upsilon, \chi_{b0}, \chi_{b1}, \ldots \} ,  
\label{Nhid}
\end{eqnarray} 
respectively, with 
\begin{equation}
\label{nx}
n_{x,y}=d_{x,y} \int \frac{\dd^3 p}{(2 \pi)^3} 
\, \exp \left(-\frac{E_{x,y}(p)}{T} \right) 
\end{equation}
the pertinent densities in Boltzmann approximation ($d_{x,y}$: 
spin-isospin degeneracy). If the open-bottom states in the QGP 
consist of $b$- and $\bar b$-quarks, Eq.~(\ref{NopH}) is replaced by
\begin{equation}
  N_{\text{op}}^b = V_{\text{FB}} \sum_{x} n_x \, , \quad x \in
  \{b, \bar b\}\ . 
\label{NopQ}
\end{equation}
As emphasized in Ref.~\cite{Goren02}, for a small number of total $b\bar
b$ pairs ($N_{b \bar b}\lsim 1$), it is quantitatively important in the
determination of $\gamma_b$ to enforce exact bottom number conservation
(\ie, in each configuration) using the canonical ensemble, which is
implemented in Eq.~(\ref{match}) by the ratio of Bessel functions. For
large values of $N_{b\bar b}$, the ratio of Bessel functions approaches
1, and one recovers the grand canonical limit.  In Eq.~(\ref{Nhid}), we
include all bottomonium states listed in Ref.~\cite{PDB04}, but due to
their relatively large masses compared to $B$-mesons (or bottom quarks),
they play no significant role in the determination of $\gamma_b$ via
Eq.~(\ref{match}).  The main uncertainty in this procedure resides in
the number, $N_{\text{op}}$, of open-bottom states, Eq.~(\ref{NopH}) or
(\ref{NopQ}), which depends on both their masses and degeneracies, as
elaborated in the following subsection. Another, more indirect
uncertainty in $N_{Y}^{\text{eq}}(T)$ pertains to the assumption of
thermal equilibrium of open-bottom states which is underlying
Eqs.~(\ref{Neq}) through (\ref{NopQ}). A pertinent (schematic)
correction will be discussed in App.~\ref{subapp_off-equil} below.

\subsection{Open-Bottom States in the QGP}
\label{subapp_open}
To assess the uncertainties associated with the open-bottom spectrum in
the QGP we will distinguish two basic scenarios (with one further
subdivision each).  The first one is that of individual $b$-quarks (with
variations in their mass), as to be expected for a weakly interacting
QGP. The second one is closer in spirit to a strongly interacting QGP,
where (colorless) hadronic correlations persist above $T_c$.  Such a
scenario has received renewed attention stimulated by findings of
hadronic bound and/or resonance states in lQCD
simulations~\cite{Karsch:2003jg,Asakawa:2002xj} (and effective
models~\cite{SZ04,MR05,Alberico:2005xw}) for $T\simeq$~1-2~$T_c$.  Also,
the strong collective phenomena (radial and elliptic flow) found
experimentally at RHIC are not easily reconciled within a perturbative
framework. On the one hand, we will therefore investigate the survival of
hadronic open-bottom states (most importantly $B$-mesons) with a varying
degree of level density toward higher masses, \ie, either $B$-meson
(and $\Lambda_b$ baryons) as listed by the particle data
group~\cite{PDB04} or additional states estimated from analogy to
open-charm hadrons (since the expected initial temperatures at LHC are
well above $2 T_c$, we consider the ``hadronic'' scenarios only for
RHIC). On the other hand, even if no hadronic states are formed in the
QGP, at moderate temperatures $T \lsim 2 T_c$, the strong coupling
constant $\alpha_s$ may not be small and induce significant
temperature-dependences in the in-medium bottom-quark mass due to
correlations with surrounding light partons.  The main point in the
following is that, at fixed $N_{b\bar b}$, differences in the
open-bottom spectrum affect the $b$-quark fugacity and thus, in turn,
modify the $Y$-equilibrium abundances. The general mechanism can be
summarized as follows~\cite{GRB04}: a larger number of available
open-bottom states (either due to a large degeneracy or a small mass
which facilitates their thermal population) implies a smaller $\gamma_b$
and thus a smaller equilibrium number of bottomonium states,
$N_Y^{\text{eq}}$.  In other words, heavier (and less) open-bottom
states favor $b$-$\bar b$ quarks to reside in bottomonia.

For definiteness, we consider the following four cases for
open-bottom states in the QGP, summarized in Fig.~\ref{fig_open}:
\begin{enumerate}
\item[(i)] 
$b$ and $\bar{b}$ quarks with a fixed mass of $m_b=5.280$~GeV, 
corresponding to the mass of the lightest $B$-meson (LHC only,
dash-dotted line in Fig.~\ref{fig_open});

\item[(ii)] $b$ and $\bar{b}$ quarks with in-medium masses $m_b(T)$,
  where the temperature dependence is inferred in analogy to
  Eq.~(\ref{mc}) as $2m_b(T) = m_{Y} + \varepsilon_B^{Y}(T)$
  (Eq.~(\ref{mb})) with the $T$-dependence of the $\Upsilon$-binding
  energy, $\varepsilon_B^{\Upsilon}(T)$, taken from Ref.~\cite{KMS88}.
  As mentioned above, here and throughout our work we assume all $Y$
  masses to be constant.  The resulting $\Upsilon$-equilibrium
  abundances are indicated by the solid (LHC) and dash-double-dotted
  line (RHIC) in Fig.~\ref{fig_open}.

\item[(iii)] open-bottom hadrons as listed in the particle data
  book~\cite{PDB04}: $\Lambda_b^0$, $B$, $B^*$, $B_s^0$, $B_c^{\pm}$
  (RHIC only, dotted line in Fig.~\ref{fig_open}).

\item[(iv)] open-bottom hadrons including states with estimated masses as 
  extrapolated from systematics of analogous (known) charmed hadrons 
  (RHIC only, dashed line in Fig.~\ref{fig_open}).
\end{enumerate}
For the two $b$-quark scenarios (i) and (ii) under LHC conditions 
(upper two curves in Fig.~\ref{fig_open}), the main difference amounts 
to a larger $\Upsilon$ equilibrium abundance for the larger $b$-quark mass
(case (i)); at high temperatures, where for case (ii) the $b$-quark
mass is essentially half the $\Upsilon$ mass (zero binding energy),
the difference is less pronounced due to the flatter thermal
distribution functions. Toward smaller temperatures (especially
close to $T_c$), $\gamma_b$ increases significantly due to a 
decreasing $N_{\text{op}}^b$, and one notices the onset of the effect 
that, in the canonical ensemble (where
$b$ and $\bar b$ have to appear together), it becomes increasingly 
favorable to have a $b\bar b$ pair in an $\Upsilon$ state which has a 
lower mass than separate $b$ and $\bar b$ quarks (the vertical decrease 
of $N_\Upsilon^{\text{eq}}$ at $T_c$ is a consequence of the volume 
expansion during the mixed phase at constant 
temperature, implying a decrease in $\gamma_b$). 

At RHIC the $b$-quark scenarios show a behavior for
$N_\Upsilon^{\text{eq}}$ analogous to LHC but with reduced magnitude
mainly due to the reduction in $N_{b\bar b}$.  Close to $T_c$, the
scenarios with hadronic open-bottom states, (iii) and (iv), lead to a
substantial increase in $N_\Upsilon^{\text{eq}}$ over the one with
in-medium $b$-quark mass, since the latter is well below the lowest
$B$-meson mass.  The sensitivity to higher-mass bottom hadrons becomes
more pronounced with increasing temperature where they are more easily
excited, thus reducing the number of bottomonium states. This, in turn,
implies that the $b$-quark scenario (i), with $m_b=5.280 {\ }
\text{GeV}=m_B$ (not shown in Fig.\ref{fig_open}), is rather similar to
the hadronic scenarios close to $T_c$.

We finally note that, during the mixed phase (\ie, at $T_c$), the 
open-bottom spectrum within each scenario is assumed not to change.
With constant temperature and increasing 3-volume, $V_{\text{FB}}(t)$, 
this implies that the bottom-quark fugacity, $\gamma_{b}(T,V_{\text{FB}})$,
decreases during the mixed phase (since $N_{b\bar b}$ is conserved), 
and with it $N_Y^{\rm eq}(T,V_{\text{FB}})$. The discontinuity in the first 
derivative of the temperature evolution (cf.~Fig.~\ref{fig_temp-profil})
at the beginning of the mixed phase is thus reflected in 
$\gamma_{b}(T,V_{\text{FB}})$ and $N_Y^{\rm eq}(T,V_{\text{FB}})$. 

\subsection{Incomplete $b$-Quark Thermalization}
\label{subapp_off-equil}

Due to the rather short duration of a few fm/c of the high-temperature
(QGP) phases in URHIC's, and a $b$-quark mass that exceeds expected
early temperatures by a large factor, $b$-quark thermalizatoin may not
be a realistic assertion. Even for charm quarks (an approach to)
thermalization appears to be unlikely under RHIC conditions if only
perturbative rescattering processes in the QGP are
considered~\cite{Svet88}. However, recent RHIC data~\cite{PHENIXv2} on
the elliptic flow of (semileptonic) decay electrons which are attributed
to $D$-mesons suggest that charm quarks undergo substantial
rescattering~\cite{Greco:2003vf} in semi-central Au-Au collisions at
$\sqrt{s}=200$~AGeV. A possible microscopic mechanism has been
identified in Refs.~\cite{HR04,Hees:2005} in terms of resonant $c$-quark
scattering with light antiquarks via $D$-meson states surviving in the
QGP at moderate temperatures, T$\lsim 2T_c$.  It was found that,
compared to perturbative calculations, the thermal relaxation times
could be reduced by up to a factor of 3-4, thus becoming comparable to
the duration of the QGP phase at RHIC.  A similar reduction was found
for bottom quarks, but with the larger mass implying larger absolute
values for the relaxation time scale by a factor of $\sim$2-3, so that
$b$-quark thermalization remains unlikely at RHIC.

Consequently, the $\Upsilon$-equilibrium abundances shown in
Fig.~\ref{fig_open} should be improved by accounting for incomplete
thermalization. As in Ref.~\cite{GR02}, we here follow a schematic
relaxation time approach, by implementing into the thermal equilibrium
abundances a reduction factor
\begin{equation}
\label{relax}
\mathcal{R} = 1 -
\mathrm{exp}\left(-\int\frac{d\tau}{\tau_{\text{eq}}}\right),
\end{equation}
where $\tau_{\text{eq}}$ is the bottom-quark equilibration time as
obtained within the resonance-scattering model~\cite{HR04}. Typical
values of $\tau_{\text{eq}}$ range from 
$\tau_{\text{eq}} \simeq 12$~fm/c at $T=300$~MeV to
$\tau_{\text{eq}} \simeq 1.1$~fm/c at $T=1$~GeV,
leading to a substantial decrease of
the regeneration term in Eq.~\eqref{rate-eq}, especially at RHIC.
We will illustrate the uncertainties induced by this correction
by considering the thermal limit in App.~\ref{app_thermal}
below.

\section{Uncertainties due to Open-Bottom States}
\label{app_open}
\begin{figure}[!tb]
\includegraphics[width=0.9\linewidth,clip=]{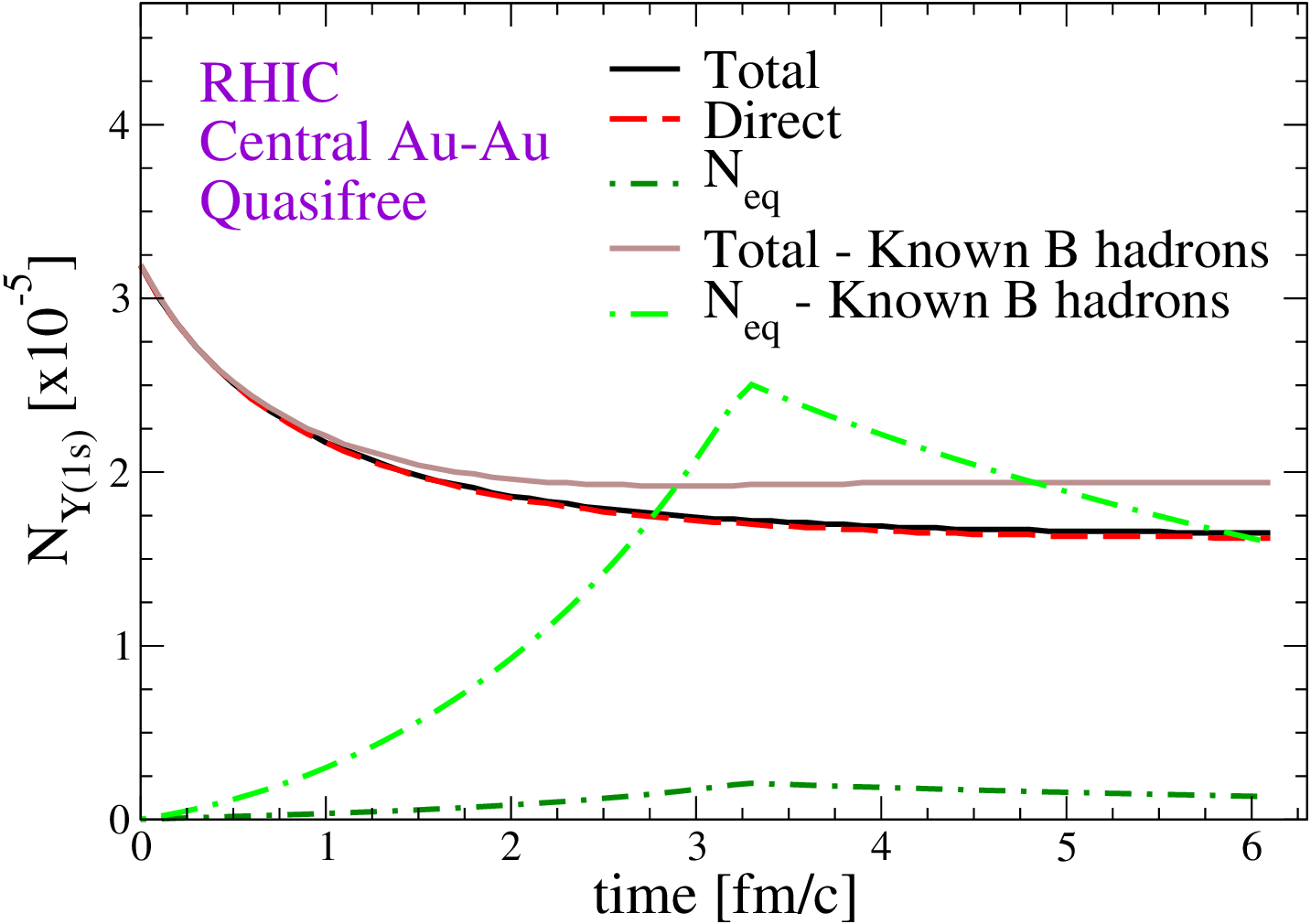}
\caption{(Color online) Time evolution of $\Upsilon(1s)$ in central
  ($b=1$~fm) Au-Au collisions at RHIC assuming different open-bottom
  spectra in the QGP. Light curves: open-bottom states constituting
  known bottom hadrons with the dash-dotted curve showing the
  $\Upsilon$-equilibrium abundance and the solid curve indicating the
  total $\Upsilon$ yield including regeneration. Dark curves:
  open-bottom states $\equiv$ in-medium $b$-quarks as in
  Fig.~\ref{fig:gluo-evol-quasi-rhic}.}
  \label{fig:regeneration-scenarios-known}
\end{figure}
\begin{figure}[!tbp]
  \includegraphics[width=0.9\linewidth,clip=]{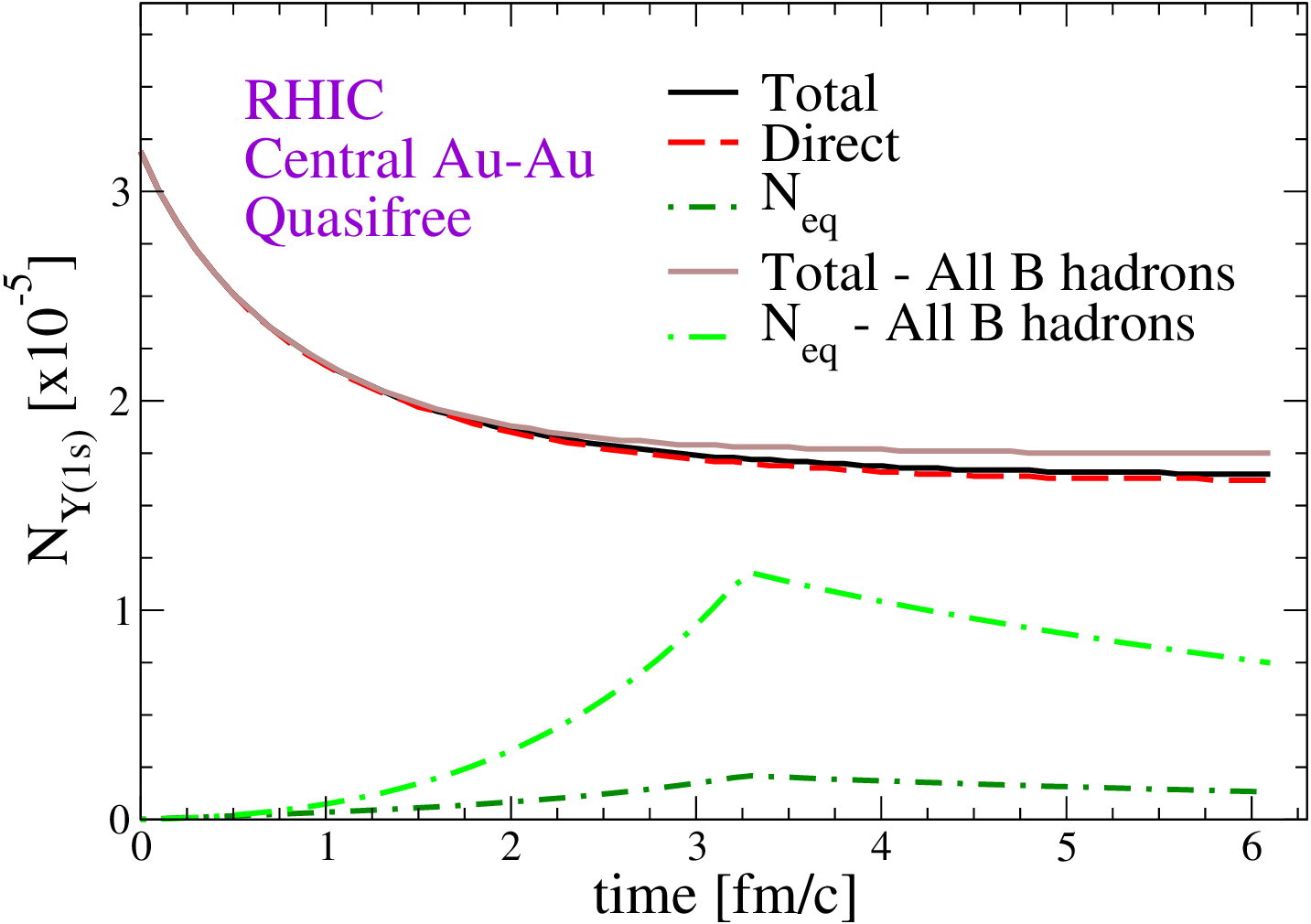}
  \caption{(Color online) Same as
    Fig.~\ref{fig:regeneration-scenarios-known} but assuming a richer
    (hadronic) open-bottom spectrum in the QGP as estimated from known
    open-charm hadrons in the vacuum.}
  \label{fig:regeneration-scenarios-all}
\end{figure}
The dependence of the $\Upsilon$-equilibrium numbers (without thermal
relaxation correction) on different assumptions about the open-bottom 
spectrum in the QGP was illustrated in Fig.~\ref{fig_open} for the 
following scenarios: 
(i) $b$-quarks with $m_b=5.280$ GeV, 
(ii) $b$-quarks with in-medium mass $m_b(T)$, 
(iii) $B$-hadrons as listed by the particle data group~\cite{PDB04},
and 
(iv) open-bottom hadrons as extrapolated from known charmed hadrons. 
In this section we study the influence of scenarios (iii) and (iv) on 
the regeneration of $\Upsilon$(1s), restricting
ourselves to RHIC energy as it seems unlikely that hadronic
correlations would persist at the temperatures anticipated 
at LHC ($T$~$\lsim$~1~GeV). We will employ the rates corresponding to
quasifree destruction with in-medium binding energies (even though
this implies an inconsistency between the rates and the $\Upsilon$ 
equilibrium number, it will allow us to estimate the maximal possible
effect as the quasifree rates are the largest considered in this work). 
Fig.~\ref{fig:regeneration-scenarios-known} pertains to the case where
the open-bottom states in the QGP are given by the known 
$B$-hadron spectrum. Since their mass tends to be larger than the
in-medium $b$-quark masses (as used in the main text, 
cf.~Fig.~\ref{fig:gluo-evol-quasi-rhic}), this scenario leads to an 
increase in the $\Upsilon$-equilibrium abundances 
(light vs. dark dash-dotted curves in 
Fig.~\ref{fig:regeneration-scenarios-known}). However, the overall
effect of a slightly enhanced regeneration is small (upper solid line).

If we assume the open-bottom spectrum to follow the same pattern as all 
known charmed hadrons, the increase in the $\Upsilon$-equilibrium number 
over the in-medium $b$-quark scenario is less 
pronounced (due to a smaller fugacity, $\gamma_b$, relative to using 
only known $B$-states), and, consequently, also the regeneration, 
leaving essentially no noticeable effect, 
cf.~Fig.~\ref{fig:regeneration-scenarios-all}.

Thus, overall, the effects of different open-bottom spectra in the QGP 
on $\Upsilon$ production at RHIC are very small.

\section{Uncertainties due to Incomplete $b$-Quark Thermalization}
\label{app_thermal}
\begin{figure}[!tbp]
  \includegraphics[width=0.9\linewidth,clip=]{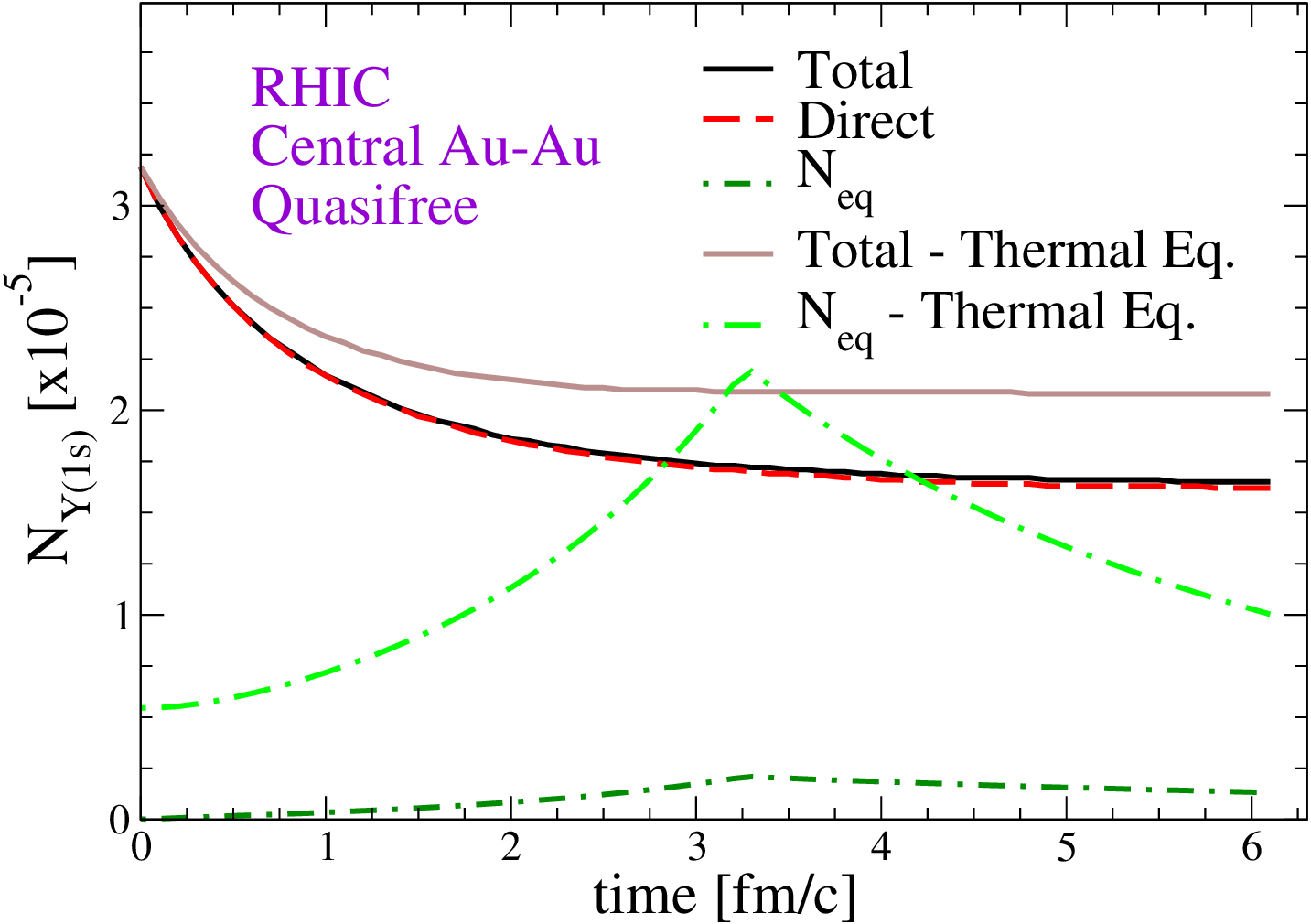}
  \caption{(Color online) Thermal off-equilibrium effects at RHIC for
    the suppression of $\Upsilon(1s)$ in central ($b=1$~fm) Au-Au
    collisions, using the quasifree process. The light set of curves
    corresponds to a full thermalization scenario with increased
    equilibrium abundances (upper dash-dotted line), leading to a 15\%
    component from secondary production (upper solid line). The darker
    lines correspond to Fig.~\ref{fig:gluo-evol-quasi-rhic} in the main
    text.}
  \label{fig:thermal-off-eq-rhic}
\end{figure}
\begin{figure}[!tb]
  \includegraphics[width=0.9\linewidth,clip=]{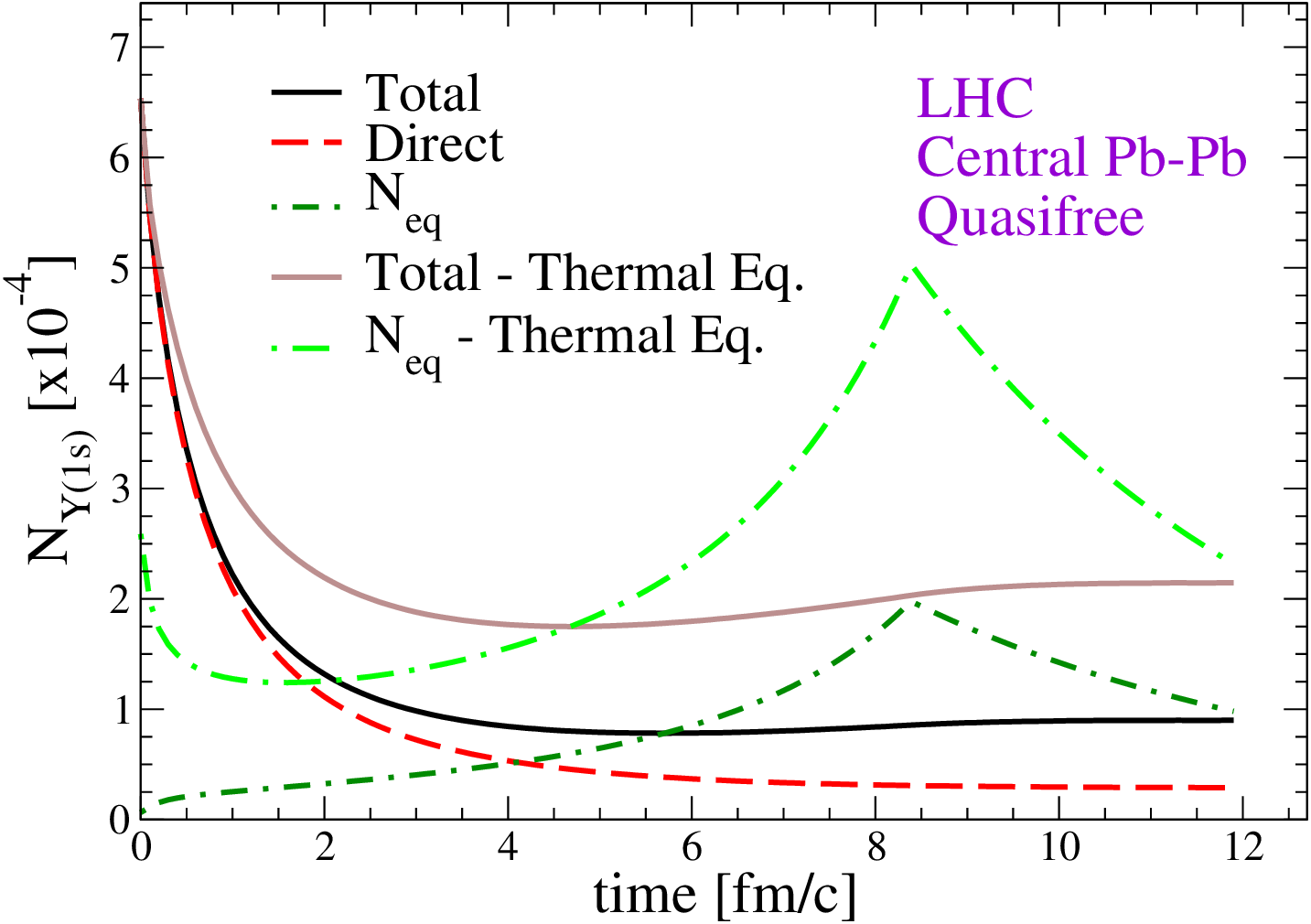}
  \caption{(Color online) Thermal off-equilibrium effects at LHC for the
    suppression of $\Upsilon(1s)$ in central ($b=1$~fm) Pb-Pb
    collisions, using the quasifree process. The light set of curves
    indicates our results assuming thermalization of the open-bottom
    states, showing larger equilibrium abundances (dash-dotted curve)
    and a larger final $\Upsilon$ yield (solid line) than with thermal
    off-equilibrium effects (dark set of curves, corresponding to
    Fig.~\ref{fig:gluo-evol-quasi-lhc} in the main text).}
  \label{fig:thermal-off-eq-lhc}
\end{figure}

Since $b$-quarks (or -mesons) are not expected to kinetically
equilibrate in central collisions at RHIC (and possibly neither at
LHC)~\cite{HR04}, all previous figures included a thermal
relaxation-time correction, Eq.~\eqref{relax}, to the equilibrium limit
of $\Upsilon$ states in the system
(Figs.~\ref{fig:gluo-evol-std-rhic}-\ref{fig:regeneration-scenarios-all}).
Due to the schematic character of its implementation, we study in this
section how our results are affected if we make the extreme assumption
of full kinetic equilibrium in the bottom sector, \ie, set ${\cal
  R}\equiv 1$. We will again do this within the quasifree dissociation
scenario including in-medium binding energies where the rates, and
therefore the sensitivity, are largest. The results are summarized in
Figs.~\ref{fig:thermal-off-eq-rhic} and \ref{fig:thermal-off-eq-lhc}.
Even though the relative increase in $N_{\Upsilon}^{\text{eq}}$ is
substantial, its impact at RHIC (Fig.~\ref{fig:thermal-off-eq-rhic}) is
still rather limited ($\sim$20\%) since it becomes comparable to initial
hard production (after nuclear absorption) only after $t\simeq2$~fm/c
when the temperature has dropped below 200~MeV and consequently the
reaction rates have become small, cf.~Fig.~\ref{fig:rates-quasifree}.
At LHC (Fig.~\ref{fig:thermal-off-eq-lhc}), 
even though the relative increase in $N_{\Upsilon}^{\text{eq}}$ is 
smaller than at RHIC (since the $b$-quark relaxation time is smaller
at larger temperatures), its absolute number is now closer to the 
actual $\Upsilon$ number, $N_\Upsilon(t)$, in the system, especially 
at intermediate times in the QGP evolution ($t\simeq$~2-4~fm/c). 
Since the reaction rates at this stage are still sizable, the final 
number in the thermal-equilibrium scenario increases by about a factor
of 2 over the off-equilibrium case. Nevertheless, the main feature of 
a strong suppression is preserved (about a factor of 3), while 
the final yield is composed to about $\sim$90\% of regenerated 
$\Upsilon$'s. This should have noticeable consequences for $p_t$-spectra 
(regenerated $\Upsilon$'s are expected to have much softer spectra than 
primordial ones, see, \eg, the recent analyses in 
Refs.~\cite{Greco:2003vf,Thews:2005vj}), and presumably even more so 
for the elliptic 
flow (compare, \eg,  Refs.~\cite{Wang:2002ck} and \cite{Greco:2003vf} 
indicating a difference of up to factor $\sim$5 in $v_2$ for the 
charmonium case).

\end{appendix}


\begin{thebibliography}{99}

\bibitem{Karsch:2003jg}
  F.~Karsch and E.~Laermann,
  In Hwa, R.C. (ed.) et al.: {\it Quark gluon plasma} 1, and
  arXiv:hep-lat/0305025.

\bibitem{Umeda:2002vr}
  T.~Umeda, K.~Nomura and H.~Matsufuru,
  Eur.\ Phys.\ J.\ C {\bf 39S1}, 9 (2005).

\bibitem{Asakawa:2003re} M.~Asakawa and T.~Hatsuda,
  Phys.\ Rev.\ Lett.\ {\bf 92}, 012001 (2004).

\bibitem{Datta:2003ww} 
  S.~Datta, F.~Karsch, P.~Petreczky and I.~Wetzorke,
  Phys.\ Rev.\ D {\bf 69}, 094507 (2004).

\bibitem{Asakawa:2002xj}
  M.~Asakawa, T.~Hatsuda and Y.~Nakahara,
  Nucl.\ Phys.\ A {\bf 715}, 863 (2003)
  [Nucl.\ Phys.\ Proc.\ Suppl.\  {\bf 119}, 481 (2003)].

\bibitem{Thews01} 
   R.L.~Thews, M.~Schroedter and J.~Rafelski,
  Phys.\ Rev.\ C {\bf 63}, 054905 (2001).

\bibitem{GRB04} 
  L.~Grandchamp, R.~Rapp and G.E.~Brown,
  Phys.\ Rev.\ Lett.\  {\bf 92}, 212301 (2004).

\bibitem{Pbm01} P.~Braun-Munzinger and J.~Stachel,
  Nucl.\ Phys.\ A {\bf 690}, 119 (2001).

\bibitem{Goren02} 
M.~I.~Gorenstein, A.~P.~Kostyuk, L.~McLerran, H.~St\"ocker and W.~Greiner,
  J.\ Phys.\ G {\bf 28}, 2151 (2002).

\bibitem{GR01} 
L.~Grandchamp and R.~Rapp,
  Phys.\ Lett.\ B {\bf 523}, 60 (2001).

\bibitem{An03} A.~Andronic, P.~Braun-Munzinger, K.~Redlich and J.~Stachel,
  Phys.\ Lett.\ B {\bf 571}, 36 (2003).
  
\bibitem{MS86} T.~Matsui and H.~Satz,
  Phys.\ Lett.\ B {\bf 178}, 416 (1986).

\bibitem{Greco:2003vf} V.~Greco, C.M.~Ko and R.~Rapp,
  Phys.\ Lett.\ B {\bf 595}, 202 (2004).
                                                                                
\bibitem{Thews:2005vj}
  R.L.~Thews and M.L.~Mangano,
  Phys.\ Rev.\ C {\bf 73}, 014904 (2006).
    
\bibitem{Svet88} B.~Svetitsky,
  Phys.\ Rev.\ D {\bf 37}, 2484 (1988).

\bibitem{Must97} M.G.~Mustafa, D.~Pal, D.K.~Srivastava and M.~Thoma,
  Phys.\ Lett.\ B {\bf 428}, 234 (1998).
 
\bibitem{HR04} H.~van Hees and R.~Rapp,
  Phys.\ Rev.\ C {\bf 71}, 034907 (2005).

\bibitem{Hees:2005}
H.~van~Hees, V.~Greco and R.~Rapp, arXiv:hep-ph/0508055.

\bibitem{GV97} 
J.F.~Gunion and R.~Vogt,
  Nucl.\ Phys.\ B {\bf 492}, 301 (1997).

\bibitem{Pal:2000zm}
  D.~Pal, B.~K.~Patra and D.K.~Srivastava,
  Eur.\ Phys.\ J.\ C {\bf 17}, 179 (2000).

\bibitem{Goncalves:2001vn}
  V.P.~Goncalves,
  Phys.\ Lett.\ B {\bf 518}, 79 (2001).

\bibitem{Peskin:1979va} M.~E.~Peskin,
  Nucl.\ Phys.\ B {\bf 156}, 365 (1979).
  
\bibitem{Bhanot:1979vb} G.~Bhanot and M.~E.~Peskin,
  Nucl.\ Phys.\ B {\bf 156}, 391 (1979).

\bibitem{YELL03} 
  M.~Bedjidian {\it et al.},
  arXiv:hep-ph/0311048.

\bibitem{LMW95} P.~Levai, B.~Muller and X.~N.~Wang,
  Phys.\ Rev.\ C {\bf 51}, 3326 (1995).

\bibitem{PDB04} 
  S.~Eidelman {\it et al.}  [Particle Data Group],
  Phys.\ Lett.\ B {\bf 592}, 1 (2004).

\bibitem{Vogt02}
R.~Vogt, Heavy Ion Phys. {\bf 18}, 11 (2003).
   
\bibitem{Vogt01} R.~Vogt  [Hard Probe Collaboration],
  Int.\ J.\ Mod.\ Phys.\ E {\bf 12}, 211 (2003).
  
\bibitem{nuc-abs}
  D.~Kharzeev, C.~Lourenco, M.~Nardi and H.~Satz,
  Z.\ Phys.\ C {\bf 74}, 307 (1997).

\bibitem{Kopeliovich:2001ee}
  B.~Kopeliovich, A.~Tarasov and J.~H\"ufner,
  Nucl.\ Phys.\ A {\bf 696}, 669 (2001).

\bibitem{GR02} L.~Grandchamp and R.~Rapp,
  Nucl.\ Phys.\ A {\bf 709}, 415 (2002).

\bibitem{Comb79}
  B.L.~Combridge, Nucl. Phys. B {\bf 151}, 429 (1979).

\bibitem{KMS88} F.~Karsch, M.T.~Mehr and H.~Satz,
  Z.\ Phys.\ C {\bf 37}, 617 (1988).

\bibitem{Alberico:2005xw}
  W.M.~Alberico, A.~Beraudo, A.~De Pace and A.~Molinari,
  arXiv:hep-ph/0507084.

\bibitem{Zhang:2002ug}
  B.~Zhang, C.M.~Ko, B.A.~Li, Z.W.~Lin and S.~Pal,
  Phys.\ Rev.\ C {\bf 65}, 054909 (2002).

\bibitem{Bratkovskaya:2003ux}
  E.L.~Bratkovskaya, W.~Cassing and H.~St\"ocker,
  Phys.\ Rev.\ C {\bf 67}, 054905 (2003).

\bibitem{RG03} 
  R.~Rapp and L.~Grandchamp,
  J.\ Phys.\ G {\bf 30}, S305 (2004).

\bibitem{Rap01} R.~Rapp,
  Phys.\ Rev.\ C {\bf 63}, 054907 (2001).

\bibitem{KR03} P.~F.~Kolb and R.~Rapp,
  Phys.\ Rev.\ C {\bf 67}, 044903 (2003).

\bibitem{Rapp:2005}
R. Rapp, 
J. Phys. {\bf G31}, S217 (2005).

\bibitem{PHENIXv2}
Y.~Akiba {\it et al.} [PHENIX Collaboration], \eprint{nucl-ex/0510008}.

\bibitem{gag05}
C.~A.~Gagliardi  [STAR Collaboration],
arXiv:nucl-ex/0512043.


\bibitem{Affolder:1999wm}  
  T.~Affolder {\it et al.}  [CDF Collaboration],
  Phys.\ Rev.\ Lett.\ {\bf 84}, 2094 (2000).

\bibitem{KLN05}
  D.~Kharzeev, E.~Levin and M.~Nardi,
  Nucl.\ Phys.\ A {\bf 747}, 609 (2005)

\bibitem{Alessandro:2004ap}
  B.~Alessandro {\it et al.}  [NA50 Collaboration],
  Eur.\ Phys.\ J.\ C {\bf 39}, 335 (2005).
    
 \bibitem{Becattini:2005hb} 
  F.~Becattini,
  Phys. Rev. Lett. {\bf 95}, 022301 (2005).
 
\bibitem{Karsch:2000kv} 
  F.~Karsch, E.~Laermann and A.~Peikert,
  Nucl.\ Phys.\ B {\bf 605}, 579 (2001).
  
\bibitem{Petreczky:2004pz} P.~Petreczky and K.~Petrov,
  Phys.\ Rev.\ D {\bf 70}, 054503 (2004).
  
\bibitem{Digal:2001ue} 
S.~Digal, P.~Petreczky and H.~Satz,
  Phys.\ Rev.\ D {\bf 64}, 094015 (2001).
  
\bibitem{Wong:2004zr} C.Y.~Wong,
  Phys.\ Rev.\ C {\bf 72}, 034906 (2005).

\bibitem{SZ04}
E.V.~Shuryak and I. Zahed, Phys. Rev. C {\bf 70}, 021901 (R) (2004).

\bibitem{MR05}
M.~Mannarelli and R.~Rapp, Phys. Rev. C {\bf 72}, 064905 (2005).

\bibitem{Umeda05}
T.~Umeda and H.~Matsufuru,
Nucl. Phys. Proc. Suppl. {\bf 140}, 547 (2005). 

\bibitem{Duraes:2002ux}
  F.O.~Duraes, H.~Kim, S.H.~Lee, F.S.~Navarra and M.~Nielsen,
  Phys.\ Rev.\ C {\bf 68}, 035208 (2003).

\bibitem{Kaczmarek:2005gi}
O.~Kaczmarek and F.~Zantow, arXiv:hep-lat/0506019.

\bibitem{Petrov:2005sd}
  K.~Petrov,
  Eur. Phys. J. C {\bf 43}, 67 (2005).

\bibitem{Back:2004je}
  B.B.~Back {\it et al.} [PHOBOS Collaboration],
  Nucl.\ Phys.\ A {\bf 757}, 28 (2005).

\bibitem{Wang:2002ck}
  X.N.~Wang and F.~Yuan,
  Phys.\ Lett.\ B {\bf 540}, 62 (2002).


\end{thebibliography}
\end{document}